\documentclass[pageno]{jpaper}
\usepackage{amsmath}
\usepackage[numbers]{natbib}
\usepackage{graphicx}
\pdfoutput=1 
\usepackage{multirow}
\usepackage{listings}
\usepackage{algorithm}
\usepackage{algorithmicx}
\usepackage[noend]{algpseudocode}
\usepackage{enumitem}
\setlist{nolistsep}
\def\code#1{{\texttt{#1}}}
\def\name#1{{\small \textsf{#1}}}

\def\policy#1{{\textsc{#1}}}

\usepackage[normalem]{ulem}

\begin{document}

\title{Preemptive Thread Block Scheduling with \\ Online Structural Runtime Prediction for Concurrent GPGPU Kernels}
\author{Sreepathi Pai \\ The University of Texas at Austin \\ sreepai@ices.utexas.edu \and R. Govindarajan \\ Indian Institute of Science \\ govind@serc.iisc.in \and Matthew J. Thazhuthaveetil \\ Indian Institute of Science \\ mjt@serc.iisc.in}

\date{25 February 2014}
\maketitle

\thispagestyle{empty}

\begin{abstract}
Recent NVIDIA Graphics Processing Units (GPUs) can execute multiple kernels concurrently.  
On these GPUs, the thread block scheduler (TBS) currently uses the FIFO policy to schedule thread blocks of concurrent kernels.  
We show that the FIFO policy leaves performance to chance, resulting in significant loss of performance and fairness.
To improve performance and fairness, we propose use of the preemptive Shortest Remaining Time First (SRTF) policy instead.  
Although SRTF requires an estimate of runtime of GPU kernels, we show that such an estimate of the runtime can be easily obtained using online profiling and exploiting a simple observation on GPU kernels' grid structure.
Specifically, we propose a novel Structural Runtime Predictor. 
Using a simple Staircase model of GPU kernel execution, we show that the runtime of a kernel can be predicted by profiling only the first few thread blocks.
We evaluate an online predictor based on this model on benchmarks from ERCBench, and find that it can estimate the actual runtime reasonably well after the execution of only a single thread block.
Next, we design a thread block scheduler that is both concurrent kernel-aware and uses this predictor.
We implement the Shortest Remaining Time First (SRTF) policy and evaluate it on two-program workloads from ERCBench.
SRTF improves STP by 1.18x and ANTT by 2.25x over FIFO.
When compared to MPMax, a state-of-the-art resource allocation policy for concurrent kernels, SRTF improves STP by 1.16x and ANTT by 1.3x.
To improve fairness, we also propose SRTF/Adaptive which controls resource usage of concurrently executing kernels to maximize fairness.
SRTF/Adaptive improves STP by 1.12x, ANTT by 2.23x and Fairness by 2.95x compared to FIFO.
Overall, our implementation of SRTF achieves system throughput to within 12.64\% of Shortest Job First (SJF, an oracle optimal scheduling policy), bridging 49\% of the gap between FIFO and SJF.

\end{abstract}

\section{Introduction}
\label{sec:intro}

Concurrent kernel execution is a relatively new and interesting feature in modern Graphics Processing Units (GPUs).
Supported by the NVIDIA Fermi and Kepler family of GPUs, it exploits task-level parallelism among independent GPU kernels. 
Unlike on the CPU, however, task level parallelism on the GPU is achieved primarily by {\it space-sharing} not {\it time-sharing}~\cite{adriaenshpca12}.
Each GPU kernel exclusively occupies the resources (i.e. registers, thread contexts, shared memory, etc.) it needs and concurrent execution is only achieved if there are enough resources left over to accommodate any concurrent kernel.
NVIDIA therefore positions concurrent kernel execution as allowing ``programs that execute a number of small kernels to utilize the whole GPU''~\cite{nvfermi}. %

Therefore, programs whose kernels already utilize the whole GPU (``large kernels'') -- the vast majority of benchmarks in Rodinia~\cite{rodinia}, Parboil2~\cite{parboil2}, etc. -- do not benefit from concurrent kernel execution and continue to execute serially~\cite{pai2013}.
However, GPU kernels are not monolithic.
Every GPU kernel is organized as a hierarchy: a {\it grid\/}\footnote{A grid is an instance of a kernel, so technically, ``concurrent grid execution'' is more accurate.} of {\it thread blocks\/}\footnote{We use CUDA terminology in this work.}.
GPU resources are actually allocated at the granularity of thread blocks, not the whole kernel.
Each thread block is also required to be independent of other thread blocks and the ordering of execution between thread blocks is not defined.
This granular execution model, designed to allow existing GPU kernels to scale and be portable across different GPU generations and models, was originally specific to CUDA~\cite{cuda} but also underpins OpenCL~\cite{opencl}.

Recent works on concurrent GPU kernel execution~\cite{adriaenshpca12,gregg2012,guevara2009,ravi2011,pai2013} have exploited this granular execution model to achieve concurrent execution for large kernels. 
They have demonstrated that controlling resource allocation of co-running kernels can improve throughput and turnaround times.
Generally, resource allocation mechanisms tackle the serialization caused by lack of resources by limiting the resources allocated to each kernel. 
In theory, kernels can always be executed concurrently because resources are always available.
However, since GPU resources are finite, these resource allocation policies only postpone eventual serialization.

This work presents {\it scheduling\/} techniques to improve concurrent kernel execution.
Orthogonal to the resource sharing policies, we show that the granular execution of GPU kernels can be exploited to obtain information that can be used to achieve better schedules for concurrent kernels.
On current GPU hardware, concurrent kernels are executed in arrival order (i.e. FIFO).
This remains the case even in the resource-allocation works cited above.
We demonstrate that FIFO is a poor choice and that preemptive scheduling policies can improve throughput and turnaround time.

This work focuses on the {\it Thread Block Scheduler\/} (TBS), the first-level hardware scheduler in GPUs~\cite{nvfermi}.
The TBS decides which thread block executes next on an execution unit (i.e. a Fermi SM or a Kepler SMX).
Once the TBS hands over a thread block to an execution unit, the second-level hardware {\it Warp Scheduler\/}, takes over.
While warp scheduling has received considerable attention~(\cite{kayiran2013,jog2013,chen2013,rhu2013}, to cite a few), TBS scheduling policies have not been examined.
Obviously, without concurrent kernels, the TBS could only choose between the thread blocks of the single kernel currently executing.
However, with concurrent kernels, the TBS can now play a significant role in improving system throughput and turnaround times.
Thus, our surprise when microbenchmarking the Fermi revealed that the TBS on the Fermi continues to issue thread blocks from concurrent kernels using a FIFO policy -- newer kernels wait until all of the thread blocks from older kernels have been issued to the SMs.
Even the latest NVIDIA GPU, the Kepler K20~\cite{keplerwp}, continues this policy.

Therefore, in this work, we demonstrate that the use of FIFO leaves performance to chance and that using appropriate preemptive scheduling policies for thread block scheduling improves both concurrent execution of kernels as well as system throughput and turnaround time. 
In particular, we propose two runtime-aware thread block scheduling policies -- Shortest Remaining Time First (SRTF) and SRTF/Adaptive -- for concurrent GPGPU workloads.
However, these policies use estimates of kernel runtime to determine their scheduling decisions when executing concurrent workloads.
We overcome this problem using an observation that the GPU kernel execution time is a simple linear function of its thread block execution time, and therefore online profiling of the execution time of first few thread blocks suffices to estimate the kernel execution time for scheduling. 
Thus, we propose a novel online runtime predictor for GPU grids to provide these estimates of runtime.
To the best of our knowledge, this is the first work to explore different TBS policies to improve performance of concurrent workloads on GPUs. 
We make the following specific contributions:
\begin{enumerate}
\item We introduce Structural Runtime Prediction and the Staircase model for online prediction of runtime of GPU kernels. This model exploits the uniform structure of grids to predict runtime.
\item We build an online runtime predictor whose predictions are within 0.48x to 1.08x of actual runtime for single-program workloads after observing only a single thread block evaluated on hardware traces.
\item Using this predictor, we implement the Shortest Remaining Time First (SRTF) policy for thread block scheduling which achieves the best system throughput (1.18x better than FIFO) and turnaround time (2.25x better than FIFO) among all policies evaluated. Our implementation of SRTF also bridges 49\% of the gap between FIFO and Shortest Job First (SJF), an optimal but unrealizable policy.
\item To improve fairness of scheduling, we propose SRTF/Adaptive, a resource-sharing and scheduling policy which ensures equitable progress for running kernels while improving STP by 1.16x, ANTT by 2.23x and Fairness by 2.95x over FIFO.
\end{enumerate}

This work is organized as follows. Section~\ref{sec:tbsmotiv} motivates the need for better thread block schedulers and online predictors. Section~\ref{sec:structural} introduces Structural Runtime Prediction and the Staircase model for prediction. In Section~\ref{sec:onlinepred} we describe the construction of an online predictor. Section~\ref{sec:tbs} describes the scheduler and scheduling policies that we evaluate. Section~\ref{sec:tbseval} evaluates our scheduler. We conclude in Section~\ref{sec:tbsconclusion}.

\section{Motivation}
\label{sec:tbsmotiv}

To evaluate the performance of the First-in First-out (FIFO) policy on concurrent kernels, we simulate the scheduling of 28 two-program workloads from the ERCBench suite~\citep{chang2010}.
For a two-program workload, FIFO's schedule is the same as either of Shortest Job First (SJF) or Longest Job First (LJF) depending on the order of arrival of the kernels.
Note that in our evaluation (Section~\ref{sec:tbseval}), there are 56 two-program workloads possible and the subset chosen arbitrarily here consists of workloads $A+B$ such that the names of benchmarks $A$ and $B$ are in alphabetical order.
In each $A+B$ workload tested, benchmark~$A$'s kernel launches before that of benchmark~$B$.

\begin{figure}
\centering
\includegraphics[scale=0.65]{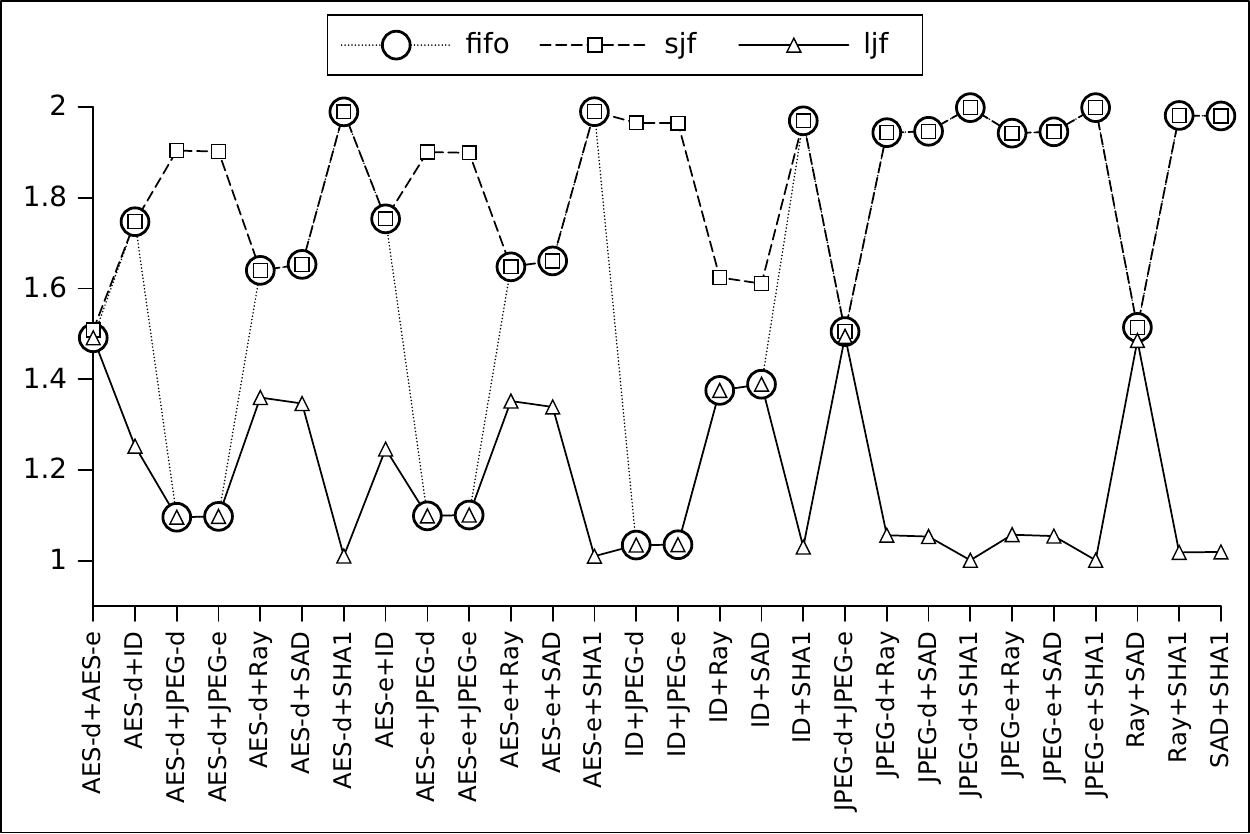}
\caption[System Throughput under the SJF, FIFO, and LJF policies]{System Throughput under the SJF, FIFO, and LJF policies for 2-kernel workloads. {\em Legend:\/} Ray=RayTracing, ID=ImageDenoising-nlm2.}
\label{fig:sjfresults}
\end{figure}

Figure~\ref{fig:sjfresults} presents the system throughput (STP, as defined in \citet{eyerman2008}) under FIFO scheduling for each of the 28 two-program workloads.
For comparison, the figure also shows the system throughput achieved by the SJF and LJF policies.
The geomean STPs are: SJF, 1.82; FIFO, 1.58; LJF, 1.16.
We observe that for 17 of the 28 workloads, FIFO achieves the same STP as SJF, that for 8 workloads its STP is the same as LJF, and for the~3 remaining workloads, the STP does not differ for SJF and LJF.
Since FIFO is oblivious to kernel characteristics, {\em these results are entirely an artefact of arrival order of the 2 kernels in each of the workloads we chose\/}. 
In this case, as kernels were launched in alphabetical order of benchmark name, in 17 of the pairs the shorter kernel started before the longer kernel.

Since NVIDIA GPUs use a FIFO policy, their throughput for concurrent kernels is also governed solely by the order in which the kernels were launched. 
For the experimental scenario described above, the Fermi would lose 15\% in system throughput on average.
In the worst case, shorter kernels will arrive while a longer kernel is already executing, so FIFO would end up scheduling like LJF and the Fermi would lose 57\% on average for these workloads.
FIFO is also non-preemptive, so execution of shorter kernels can end up being serialized behind those of larger kernels.
This serialization at the GPU level can lead to slowdowns of the CPU part of the program as well.
For example, TimeGraph~\citep{kato2011} demonstrates that GPU programs with high OS priorities can suffer from priority inversion when the GPU is monopolized by a long-running kernel from a lower-priority program.

Recently proposed resource reservation policies~\citep{adriaenshpca12,gregg2012,guevara2009,ravi2011,pai2013} partially address the problem by reserving resources for every kernel that is running but continue to use FIFO.
While this  prevents serialization by guaranteeing access to the GPU, our evaluation of a state-of-the-art reservation policy will show (Section~\ref{sec:tbseval}) that policies other than FIFO can lead to better performance.
In particular, since thread blocks can reordered without violating CUDA semantics, a TBS can make the following decisions if a new grid arrives while an old grid is executing:
\begin{enumerate}
\item {\it Do nothing:\/} Continue executing thread blocks from the currently running grid. 
\item {\it Run with available resources:\/} Issue all thread blocks from the currently running grid and if resources are leftover, attempt to schedule thread blocks from any concurrently running grid. As resources on an SM are allocated at the granularity of an entire thread block, some grids may underutilize resources potentially permitting their use by concurrently running grids.
\item {\it Adjust grid resources:\/} Vary the number of thread blocks or the number of threads in a thread block~\cite{pai2013}, in order to distribute a SM's resources between concurrently executing grids.
\item {\it Preempt running grid:\/} Pause scheduling of thread blocks from the current grid, while allowing thread blocks from other grids to be scheduled in order to prevent serialization of short kernels or enforce OS priorities.
\end{enumerate}

The first item in the list above describes FIFO execution.
Past resource-sharing policies can be described (~\cite{adriaenshpca12,gregg2012,guevara2009,ravi2011,pai2013}, by the second and third items.
In this work, we primarily focus on the fourth item, i.e. switching between grids, but also describe a dynamic grid resource adjustment policy.

To implement an SJF-like scheduling policy, a TBS requires knowledge of the runtimes of currently executing and the newly arrived grids.
There are two main techniques that could be used to obtain runtimes in advance.
Offline models~\cite{sim2012,baghsorkhi2010,kothapalli2009} could be used to predict runtimes.
Alternatively, a historical database of runtimes could be maintained per kernel \cite{luk2009,jia2012,gregg2011,diamos2008,belviranli2013,jimenez2009} and used to predict runtimes.

The primary disadvantage of offline models is that they are built for specific GPUs and require profile information for every kernel that may run.
This is impractical in general.
Predictors that use historical databases fare better since they use profile information, but they are unable to make predictions until they have seen enough {\it complete\/} runs.
Crucially, since none of these predictors handles concurrent kernels at all, they cannot be used when scheduling thread blocks of concurrent kernels.

Ideally, an online predictor that is both aware of concurrent kernels and that can predict runtime for all kernels in advance -- either at kernel launch or after a few thread blocks have finished executing -- is needed.
Such a predictor could be used by the TBS to make scheduling decisions.
Therefore, this work develops: (i) an online runtime predictor for GPU kernels, and (ii) thread block scheduler policies that use this predictor.

\section{Structural Runtime Prediction}
\label{sec:structural}
We introduce the principle of {\it Structural Prediction\/} on which our online predictor is based.
Structural prediction essentially treats the execution of a grid's $N$ thread blocks (all of which have the same code) as $N$ repeated executions of the same program.
So by profiling the first few thread blocks of a grid, we can predict the behaviour of the remaining thread blocks.
In this work, we observe the runtime of a thread block and use it to predict the runtime of the whole kernel.
We call this technique {\it Structural Runtime Prediction\/}.

\subsection{The Staircase Model}

When a grid is launched on an NVIDIA Fermi\footnote{To avoid clumsy sentence constructions, we refer exclusively to the Fermi. However, the results and observations in this section are valid on the Kepler as well. See supplementary Section~\ref{sec:kepler} for details.}, its thread blocks are mapped to one of the Fermi's many streaming multiprocessors.
Each SM has a finite number of resources (registers, threads, shared memory, block contexts) that are allocated at thread block granularity.
A SM accommodates as many thread blocks of the grid as possible until one of these resources runs out.
The maximum number of thread blocks of a grid that can be accommodated on an SM is called the {\it maximum residency\/} of that grid. %
Thread blocks that cannot be accommodated wait in queue until resources are available.
When a running thread block finishes, the Fermi thread block scheduler schedules a queued thread block in its place.
The grid has finished executing when all its thread blocks are done.

\begin{figure}
\centering
\includegraphics[scale=0.65]{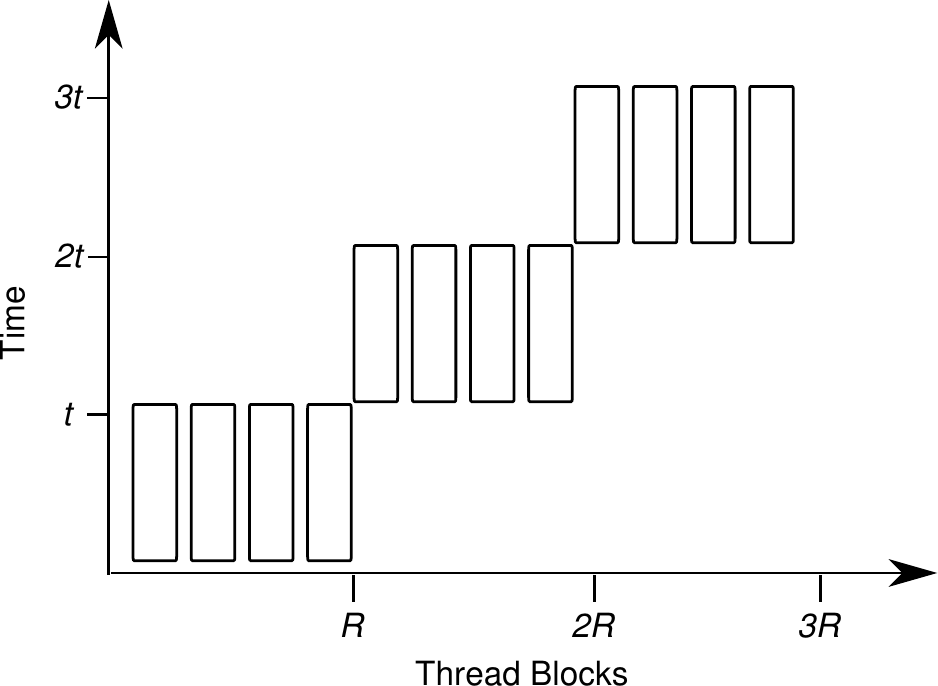}
\caption[Staircase Model Execution of $N$ thread blocks on a single SM]{Staircase Model Execution of $N$ thread blocks on a single SM, with maximum residency $R = 4$ and each block taking $t$ time to execute. Here, $N = 3R$.}
\label{fig:model}
\end{figure}

Figure~\ref{fig:model} illustrates this model execution of a grid on a single SM.
This grid has a maximum residency of $R=4$ and each thread block executes in $t$ time.
From the figure, therefore, the total time for execution of all $N$ blocks assigned to a single SM is therefore a simple function:
\begin{equation}
T = (\lceil{N / R}\rceil)\,t
\label{eqn:total}
\end{equation}

If we assume an even distribution of $B$ thread blocks across $N_{SM}$ SMs, then $N = B / N_{SM}$.
The maximum residency $R$ can be determined at grid launch time using using formulae like those in the NVIDIA Occupancy Calculator~\citep{cudaocc}.
Then, to predict runtime, Equation~\ref{eqn:total} only needs the value of $t$.
This could be obtained by sampling, possibly as soon as a single thread block finishes execution.
However, for the prediction to be useful for scheduling, the grid must execute more than $R$ blocks per SM as otherwise predictions are not timely.

\subsection{Staircase Model Evaluation}
\label{sec:smodeleval}
To evaluate the Staircase model, we instrument major kernels in the Parboil2~\citep{parboil2} and the ERCBench~\citep{chang2010} suites to record the start and end time of each thread block and the SM it was executed on.
We run these instrumented kernels on a Fermi-based NVIDIA Tesla C2070, with a quad-core Intel Xeon W3550 CPU, 16GB RAM and running Debian Linux 6.0 (64-bit) with CUDA driver~295.41 and CUDA runtime~4.2.

\begin{figure}
\centering
\includegraphics[scale=0.45]{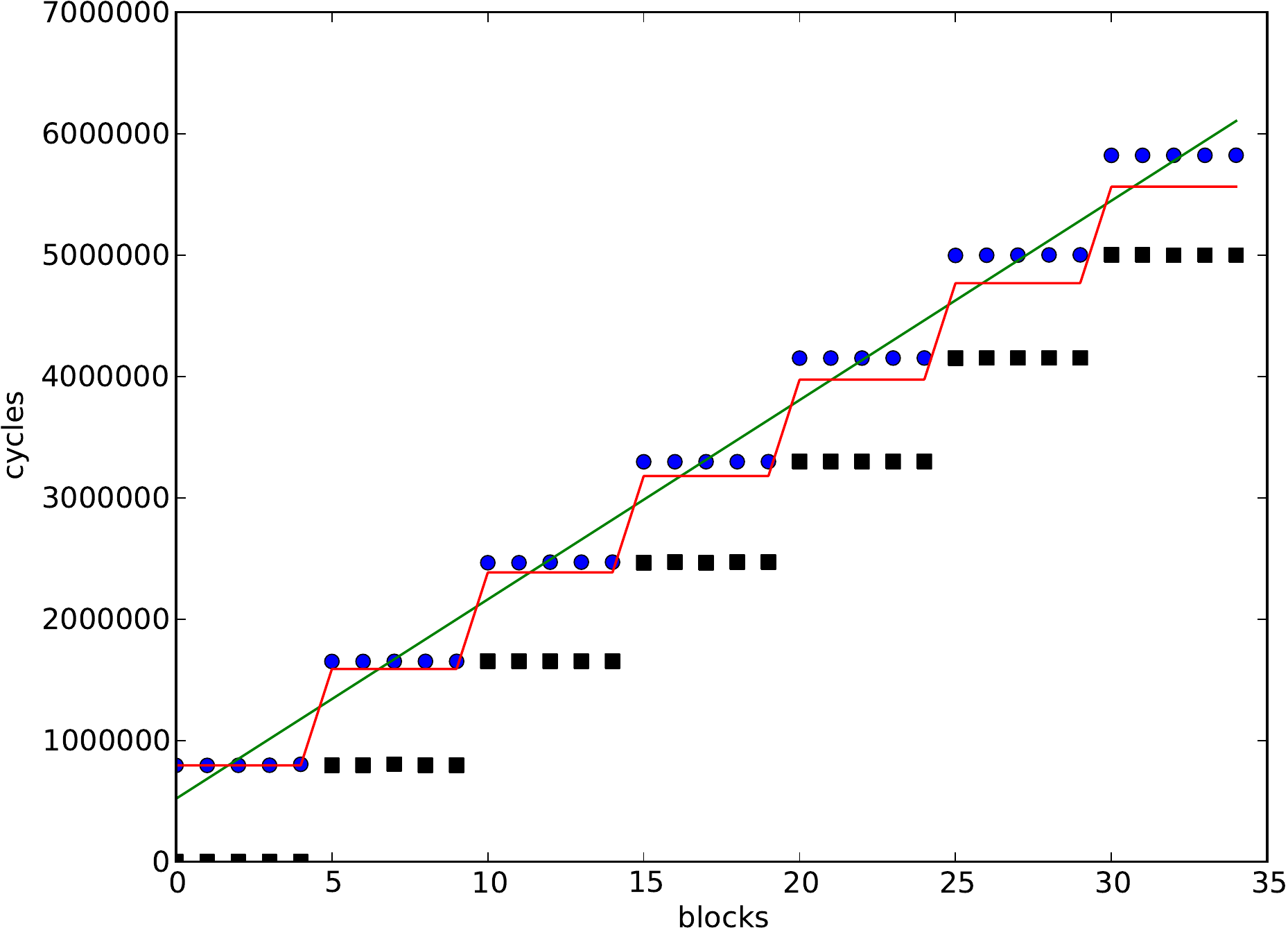}
\caption[Execution of SGEMM's thread blocks from one SM]{Execution of SGEMM's thread blocks on one SM. Blocks are ordered by finishing time. Black squares represent start times of each thread block, dark blue circles denote ending time. Green line is linear fit to all the end timings. Red line is prediction from equation~\ref{eqn:total}, with $t$ being the duration of the first block to finish.}
\label{fig:sgemm}
\end{figure}

Figure~\ref{fig:sgemm} plots the end times of the thread blocks of the Parboil2 SGEMM kernel from a single SM.
This instance of SGEMM execution closely resembles the Staircase model execution of~Figure~\ref{fig:model}.
Also shown is the linear fit to these end times using least-squares linear regression, as well as the runtime value ($T$) predicted using Equation~\ref{eqn:total} with $t$ set to the duration of the first finishing block.
The linear fit overestimates the actual finish time by 4.8\% while the staircase model prediction underestimates it by 6.04\%.

\begin{figure}
\centering
\includegraphics[scale=0.45]{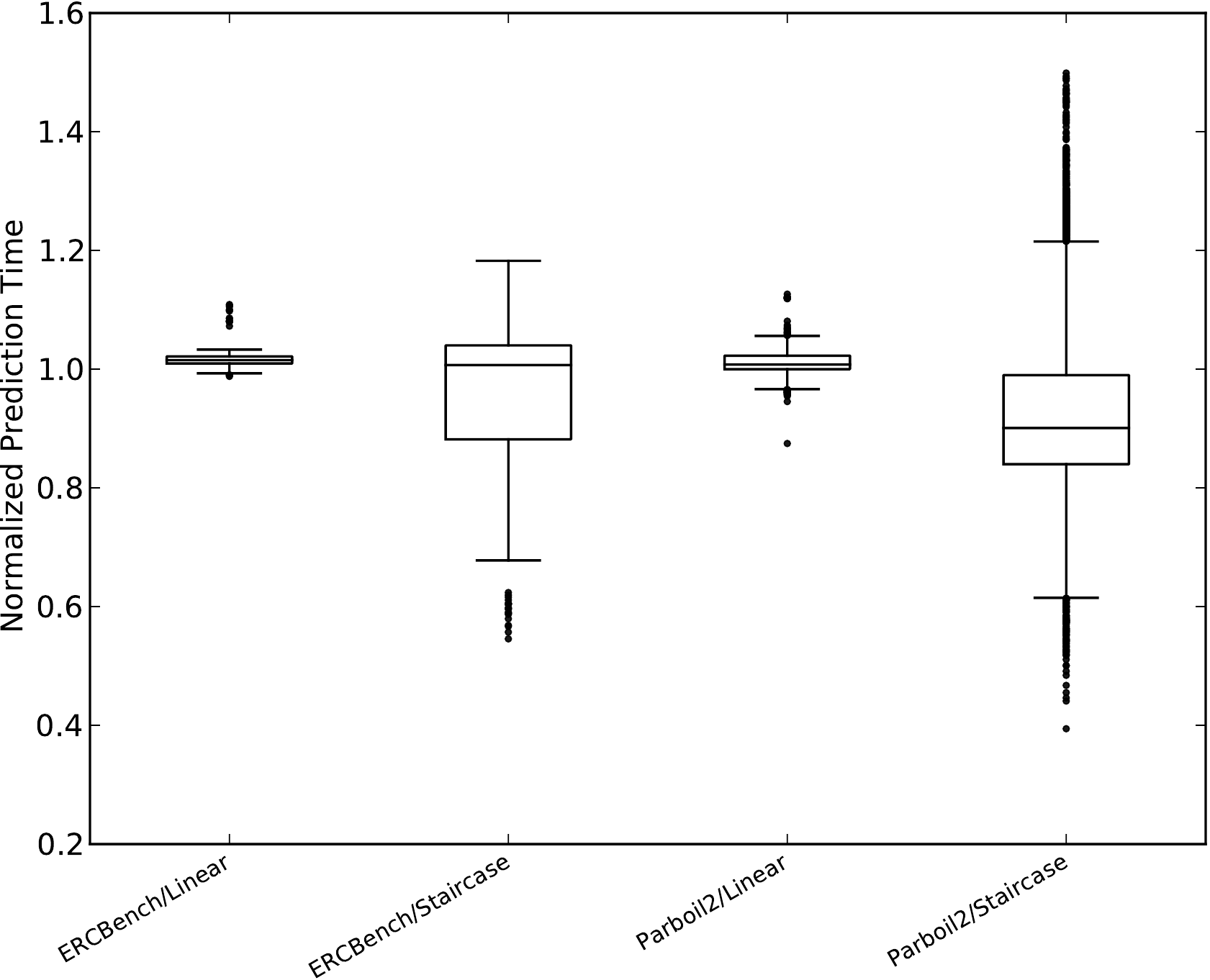}
\caption[Boxplots of Predictions from Linear Regression and Staircase Models]{Boxplots of Predictions from Linear Regression and Staircase Models for ERCBench and Parboil2 benchmarks normalized to actual runtime. }
\label{fig:linear_model}
\end{figure}

We now obtain predictions for the other kernels.
Predictions are obtained for every invocation of a kernel and on every SM.
Since the Fermi has 14 SMs, and some kernels are invoked multiple times, we obtain 4522 predictions for Parboil2 kernels and 112 predictions for the kernels in ERCBench.
Figure~\ref{fig:linear_model} is a boxplot of predictions normalized to the actual runtime obtained using both linear regression and Equation~\ref{eqn:total} for ERCBench and Parboil2.
Outliers (lying beyond the 1.5 inter-quartile range) are also plotted.
Linear regression results in normalized predictions between 0.99x to 1.11x of actual runtime for ERCBench and 0.87x to 1.13x for Parboil2.
This strongly supports our hypothesis that GPU kernel runtime is a linear function. 

Unlike the models constructed by linear regression which are built using the end times of all thread blocks, predictions from Equation~\ref{eqn:total} only use the duration of the first thread block.
These predictions normalized to actual runtime lie between 0.54x to 1.18x for ERCBench and 0.39x to 1.49x for Parboil2.
If we exclude outliers, normalized predictions are between 0.66x and 1.18x for ERCBench and 0.6x and 1.2x for Parboil2.
We investigate the major causes for this inaccuracy in the following sections.

\subsection{Non-Staircase Model Behaviour}
\label{sec:nonstaircase}
\begin{figure}
\centering
\includegraphics[scale=0.45]{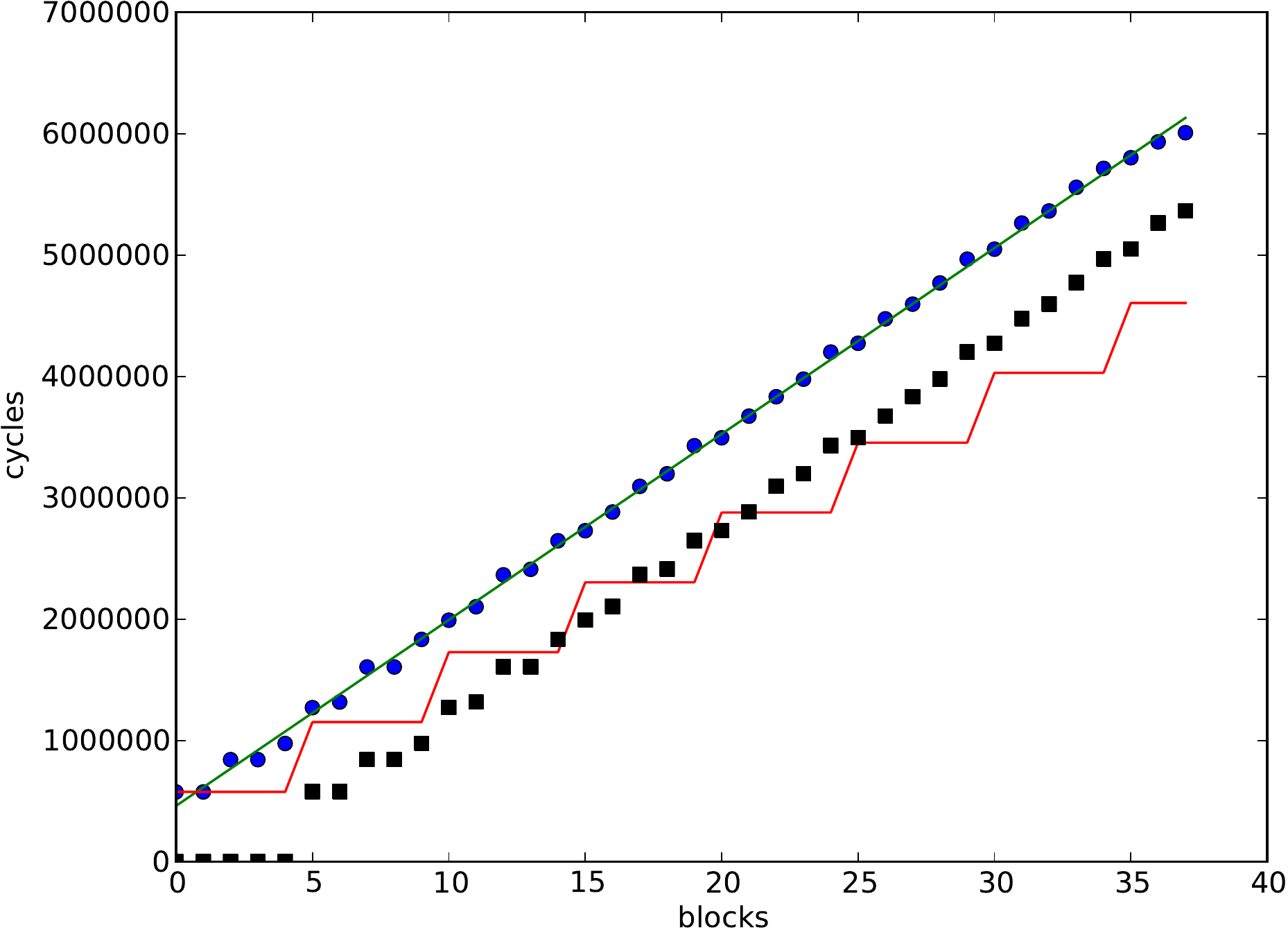}
\caption[SGEMM execution as recorded on a different SM]{SGEMM execution as recorded on a different SM from that of Figure~\ref{fig:sgemm}. Start times
  of all subsequent blocks are staggered by the end times of the first 5
  blocks. Again, black square represent start times of a thread block, dark blue circles
  represent ending times, the green line depicts the linear fit, and
  the red line depicts the value of equation~\ref{eqn:total}.}
\label{fig:sgemm_inacc}
\end{figure}

Figure~\ref{fig:sgemm_inacc} presents a case where prediction using Equation~\ref{eqn:total} underestimates the total runtime.
In the figure, from the same execution as Figure~\ref{fig:sgemm} but from a different SM, the first $R$ blocks each end at different times.
As a result, the starting times of subsequent blocks are staggered.
While the runtime continues to be linear, direct application of Equation~\ref{eqn:total} to such executions leads to gross underestimates.
In ERCBench, all underestimates (< 0.9x) were due to staggered execution affecting executions of AES-d and SHA1 on some SMs.
We observe such staggered executions commonly on hardware, but they were entirely absent in the simulator that we used.

\subsection{Systematic Variations in Duration per Thread Block}
\label{sec:tvariations}
The duration of each thread block, $t$, can vary during execution due to both random errors and systematic factors.
In this section, we look at the factors that systematically affect $t$ and therefore need to be accounted for in the prediction mechanism.

\subsubsection{Differing Work per Thread Block}

Although the CUDA Programming Guide~\citep{cuda4} recommends that work per thread block be uniform for best performance, it does not require it, and thus $t$ can be non-uniform across thread blocks.
For example, if the work done by thread blocks in the kernel differs based on the {\it value\/} of their inputs, then each thread block could take a different amount of time to run.
Even if all blocks were written to perform the same amount of work, we observe a major source of overestimation error (> 1.1x) from startup effects in the first few thread blocks whose longer than average duration leads to overestimates.
In ERCBench, some SMs executing JPEG-d, SAD and SHA1 exhibit this behaviour.

\begin{figure}
\centering
\includegraphics[scale=0.45]{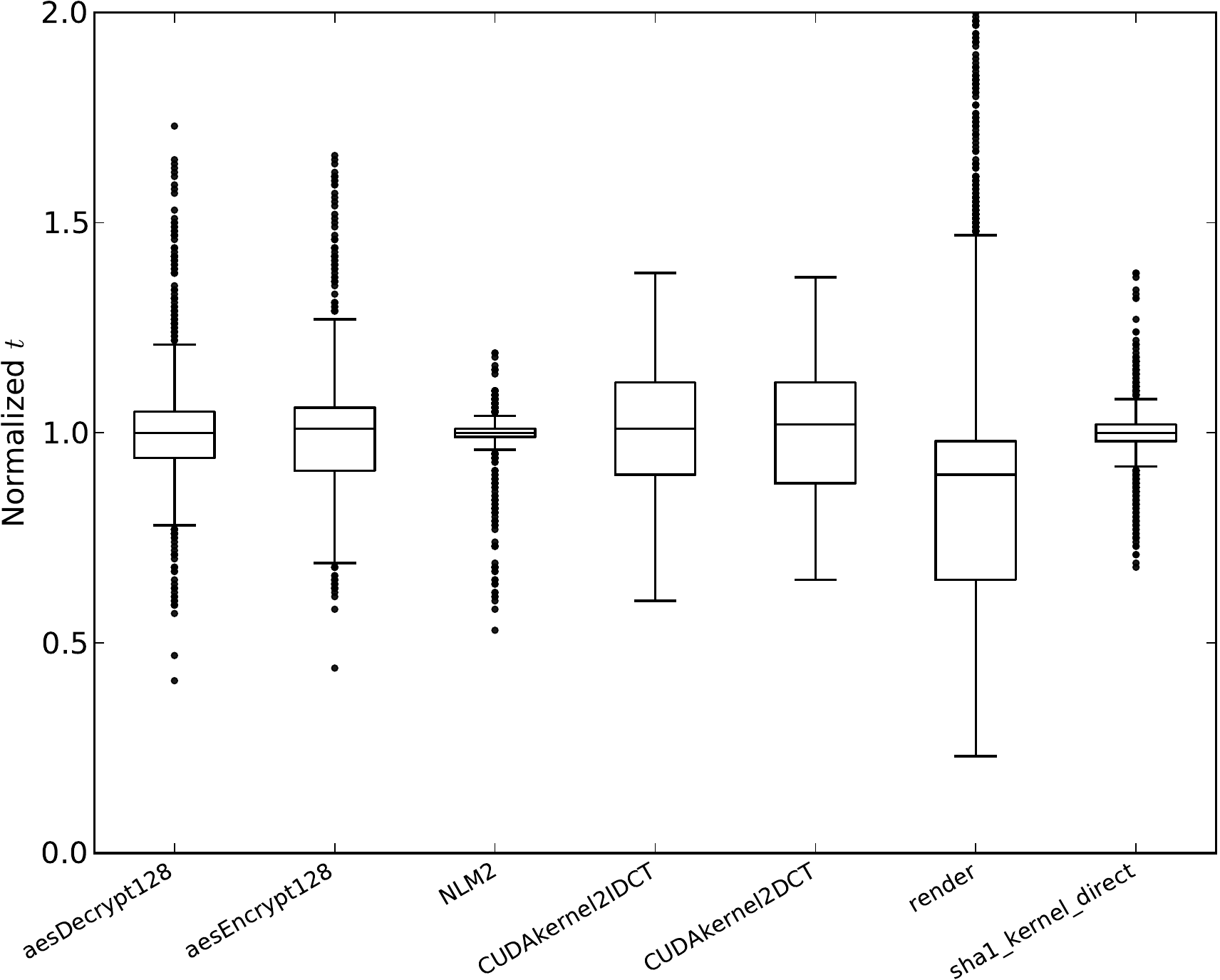}
\caption[Boxplots of normalized thread block durations]{Boxplots of thread block durations ($t$) normalized to their average for a kernel. \code{render}'s maximum value is 4.}
\label{fig:blocktimespread}
\end{figure}

We use the data from Section~\ref{sec:smodeleval} to examine the distribution of thread block durations.
Figure~\ref{fig:blocktimespread} shows the boxplot of thread block durations ($t$) normalized to their average for kernels in the ECRBench suite.
Observe that values of $t$ for the majority of thread blocks are within 0.95x to 1.1x of the average except in the case of RayTrace's \code{render} kernel.
This is expected since \code{render}'s thread blocks perform differing amounts of work. 
But even in this case, we can see 50\% of its thread blocks are within 0.75x to 1x of the mean.
Furthermore, despite the magnitude of deviation, the linear regression model for RayTrace predicts with a maximum error of 9\%, while even Equation~\ref{eqn:total} has a maximum error of 18\%.
For Parboil2 (not shown here), the major long-running kernels tend to have uniform thread block durations, but the smaller kernels can exhibit non-uniformity -- \code{cutcp}'s thread block times vary from 0.4x to 1.37x of the average.
We conclude that the majority of kernels demonstrate the tendency to perform nearly uniform amount of work per thread block.
For those kernels that do not, our predictor implementation~(Section~\ref{sec:onlinepred}) uses the actual runtime as feedback to correct any drift.

\subsubsection{Differing SM Behaviour}

Figures~\ref{fig:sgemm} and~\ref{fig:sgemm_inacc} demonstrate that individual SMs can vary in their behaviour for the same kernel during the same run.
Some GPU programs, such as those studied by \citet{liu2009} and \citet{samadi2012}, exhibit load imbalance across SMs when {\it sizes\/} of their inputs are varied.
To obtain reliable predictions for these programs we implement per-SM predictors.

\subsubsection{Effect of Residency}

\begin{figure}
\centering
\includegraphics[scale=0.35]{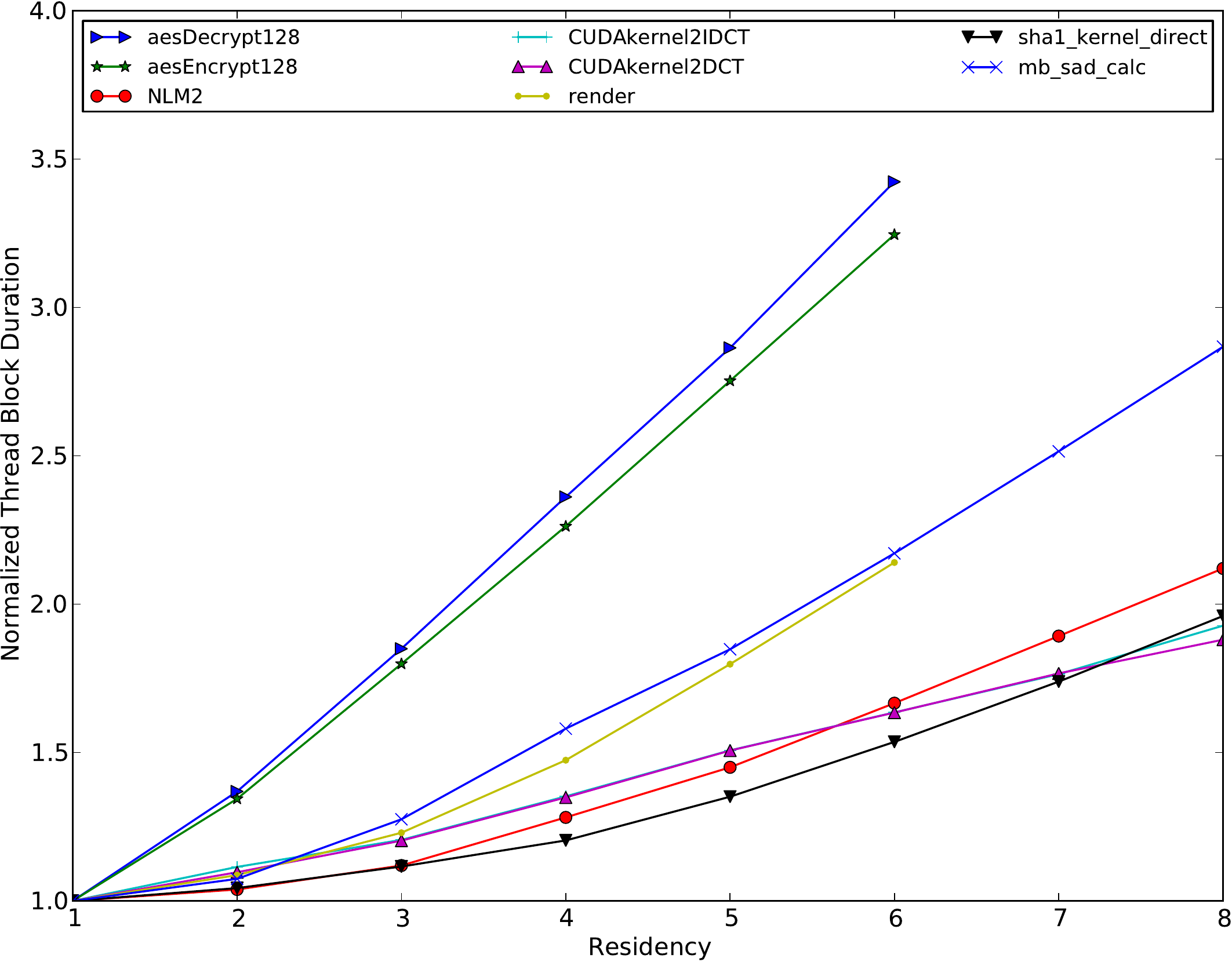}
\caption[Average thread block duration at various residencies]{Average thread block duration at various residencies normalized to average thread block duration at residency 1. AES-e, AES-d and render have a maximum residency of 6 blocks and all other kernels have maximum residency of 8 thread blocks.}
\label{fig:tbtimes_residency}
\end{figure}

\begin{figure}
\centering
\includegraphics[scale=0.35]{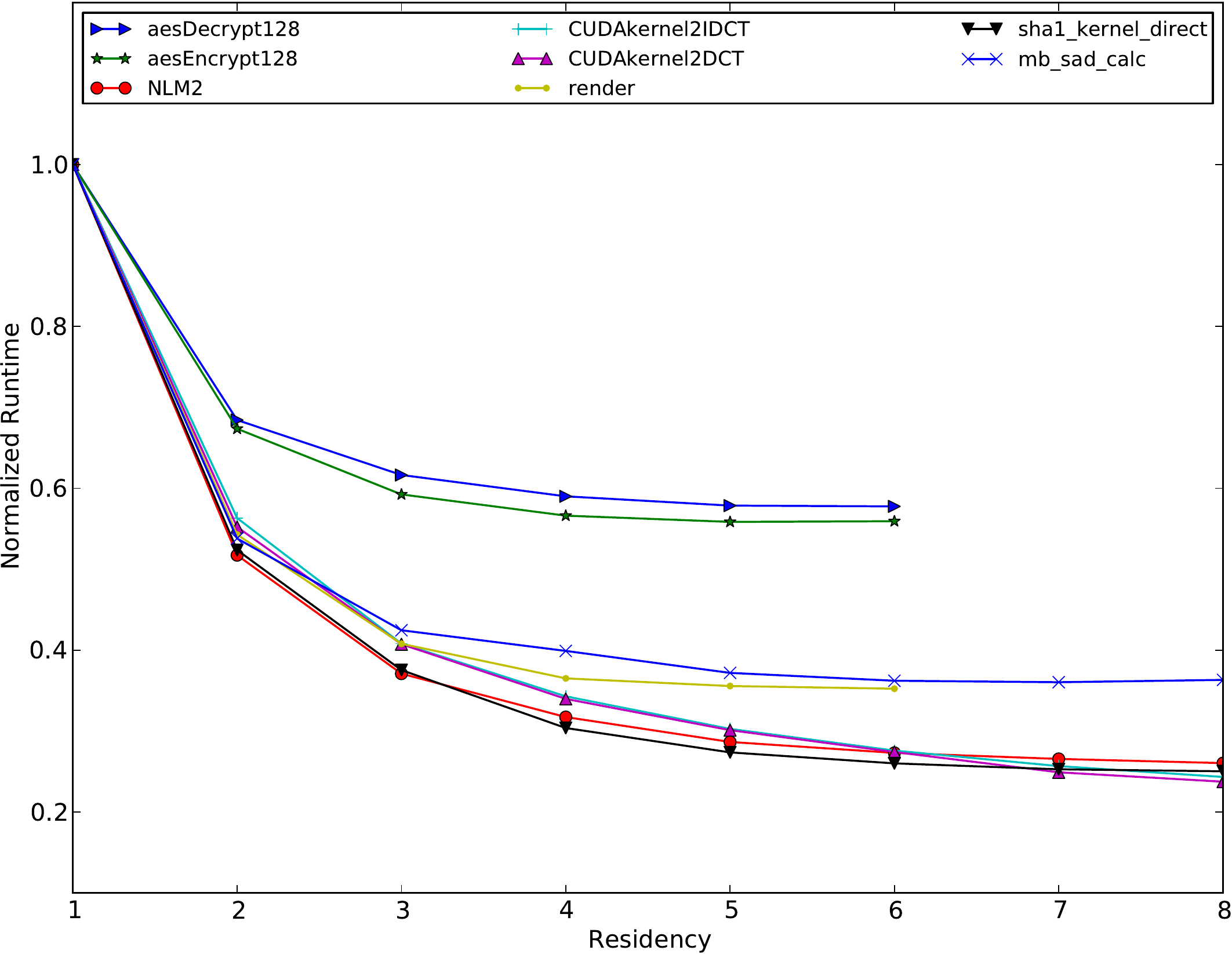}
\caption[Total kernel runtime at various residencies]{Total kernel runtime at various residencies normalized to runtime at residency 1.}
\label{fig:runtimes_residency}
\end{figure}

Although each kernel has a fixed maximum residency $R$, non-availability of resources during concurrent kernel execution might limit the number of resident thread blocks.
Therefore, we investigate the effects of residency on $t$.
To separate out the effects of co-runners, these experiments are run on hardware with each kernel running alone at different residencies.
The next section considers the effects of co-runners.
Figure~\ref{fig:tbtimes_residency} shows the variation in $t$ as residency is varied for a kernel.
The values of $t$ are smallest when residency is 1 and increase with residency.
However, as residency is increased, Figure~\ref{fig:runtimes_residency} shows that {\it total runtime\/} decreases and ultimately saturates.
Thus, increases in $t$ are offset by the increase in throughput due to increased residency.

The actual rate of increase in $t$  for a kernel as the residency increases is a non-linear function and depends on both the kernel and the GPU it is running on.
For example, \name{SHA1} has a maximum residency of 8 thread blocks.
However, at 64 threads to each thread block, there are only 16 warps at maximum residency.
Therefore, on the Fermi which can issue two instructions per clock, \name{SHA1} is unable to supply enough instructions to saturate issue at low residencies ($<4$). 
However, once its residency increases beyond 4, it supplies at least two instructions per cycle, but its performance is now limited by two other factors: shared memory bandwidth, and its limited ILP due to long dependent chains of consecutive instructions.
We leave the detailed modeling of these interactions to future work.
In our predictor, we simply resample $t$ whenever residency changes.

\subsubsection{Effects of Co-runners}
\label{sec:corunners}
The effect of co-runners on $t$ was studied by running thread blocks from different kernels together.
Unlike the data presented up to this point in this paper, the results in this section are necessarily from the simulator (Section~\ref{sec:simulator}).

\begin{figure}
\centering
\includegraphics[scale=0.35]{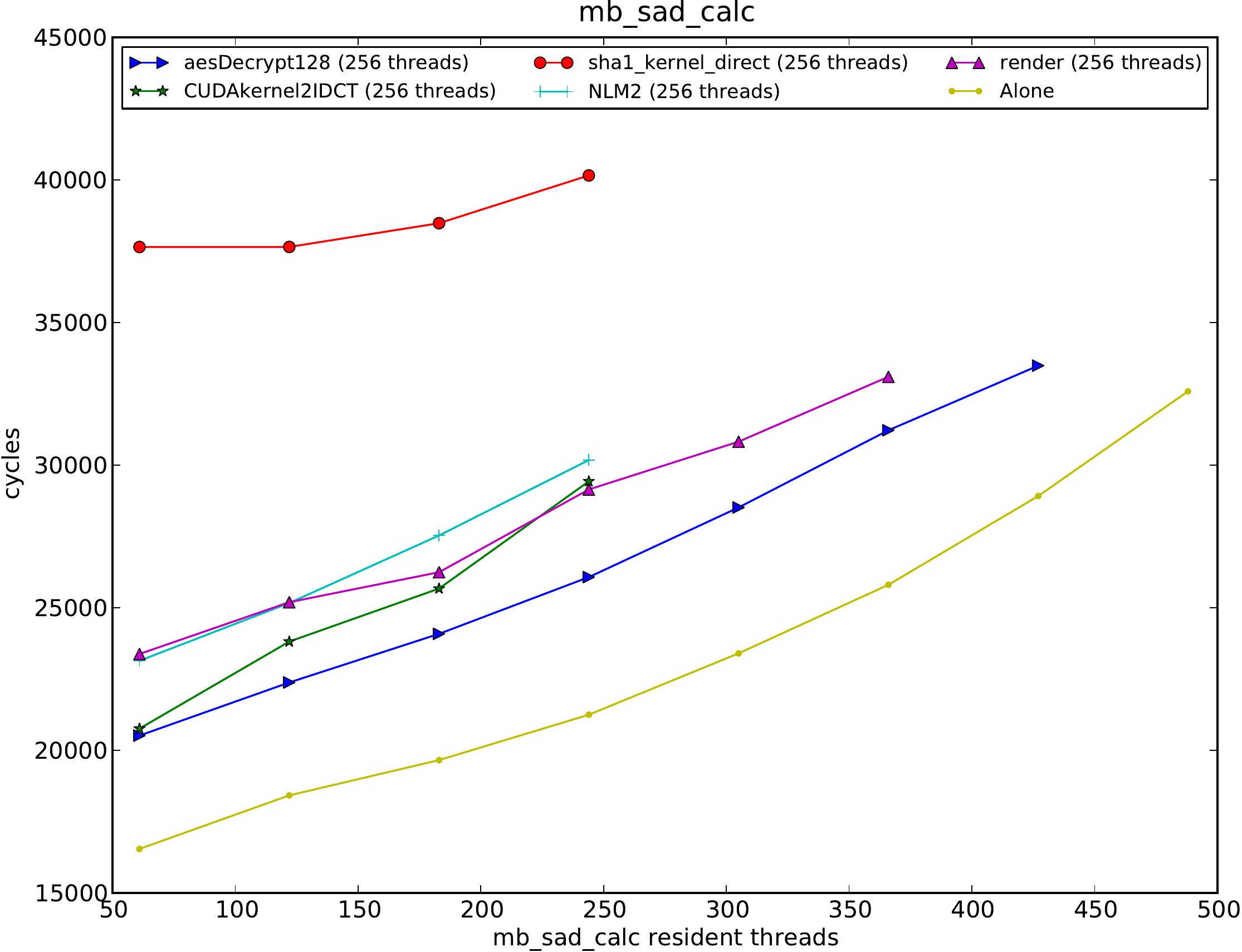}
\caption[Average thread block durations from simulator for SAD mb\_sad\_calc kernel at different residencies when running with different kernels]{Average duration of a thread block from simulator for SAD mb\_sad\_calc kernel (61 threads per block) at different residencies when sharing the GPU with 256 threads of a co-running kernel. }
\label{fig:tbtimes_sharing}
\end{figure}

Figure~\ref{fig:tbtimes_sharing} shows the effect on \code{mb\_sad\_calc}'s average thread block durations at different residencies.
In this experiment, \code{mb\_sad\-\_calc} runs along with 256 threads of another kernel.
From the figure, we observe that the effect on average thread block duration varies depending on the co-running kernel.
The 256 threads of \name{SHA1} result in a greater change in the thread block duration of \code{mb\_sad\_calc} than any other kernel.

\begin{figure}
\centering
\includegraphics[scale=0.35]{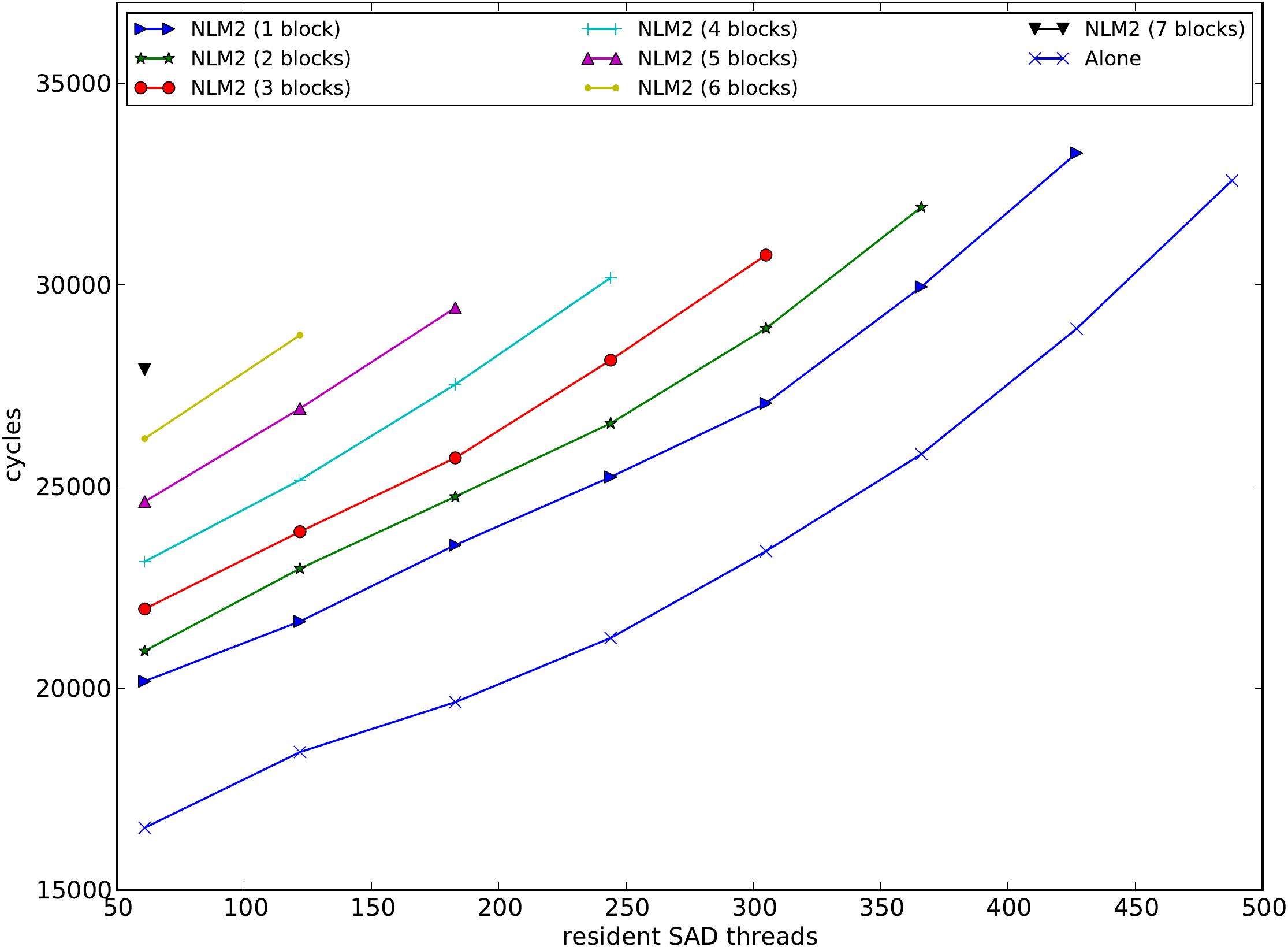}
\caption[Average thread block durations from simulator for SAD mb\_sad\_calc kernel when running with NLM2]{Average thread block durations from simulator for SAD mb\_sad\_calc kernel (61 threads per block) when sharing the GPU with NLM2 as residencies are varied for both kernels. }
\label{fig:tbtimes_sharing_other_kernel}
\end{figure}

Figure~\ref{fig:tbtimes_sharing_other_kernel} shows the effect on the thread block duration of \code{mb\_sad\_calc} as the number of threads of co-running \code{NLM2} are varied.
The duration per thread block for \code{mb\_sad\_calc} varies from approximately 16000 cycles when alone to nearly 28000 cycles when running with seven thread blocks of \code{NLM2}.

Clearly, runtime is affected by both the partitioning of resources and the specific co-running kernel.
In our predictor, therefore, we resample  $t$ whenever the set of co-runners or their residencies changes.

\subsection{Summary}
\label{sec:structconcl}
The runtime of a GPU kernel running alone at a fixed residency can be modeled as a linear function.
While the models obtained by linear regression are accurate, they must be built using data from complete runs as using a limited number of thread blocks affects accuracy.
We found that linear models built using the end times of the first $R$ thread blocks were not very accurate.
Linear models that used the first $2R$ thread blocks fared better, predicting between 0.8x to 2x of actual runtime for the ERCBench kernels.
However, in terms of runtime, $2R$ blocks represent 7\% to 65\% of total ERCBench kernel runtimes (median 18\%), thus compromising on timeliness.

Runtimes of concurrently running kernels, on the other hand, are at best only {\it piecewise linear\/}, necessitating frequent resampling of $t$.
Equation~\ref{eqn:total}, which only requires the duration of a single thread block is therefore a better choice to predict concurrent kernel runtime as it can be extended easily to deal with them (Section~\ref{sec:onlinepred}).
In fact, we find that for the scheduling policies we evaluate, a rough but early prediction is just as useful as an accurate oracle-supplied prediction (Section~\ref{sec:detailedstp}).

\section{The Simple Slicing Predictor}
\label{sec:onlinepred}

The Simple Slicing (SS) runtime predictor is an online, concurrent-kernel aware predictor based on Equation~\ref{eqn:total} that takes Sections~\ref{sec:nonstaircase}--\ref{sec:structconcl} into consideration.
Its prediction of runtime is an estimate of how much time a kernel would take to complete if it was running from now (i.e. the time at which the prediction is made) to completion, under the current conditions ($t$, residency and co-runners).

To accommodate changes in $t$ as the kernel executes (Section~\ref{sec:tvariations}--\ref{sec:corunners}), we split the execution of a kernel into multiple {\it slices\/}.
Each slice is demarcated by any of the events that cause changes in $t$.
In our current design, kernel launches and kernel endings mark the boundaries of slices for all running kernels.
We assume that $t$ remains constant within a slice enabling the predictor to predict timings for blocks in that slice.
Our predictions assume that the last thread block to execute is contained in the current slice since slice boundaries cannot be predicted in advance.
Finally, as each SM can vary in behaviour, our predictor predicts runtimes for each kernel on a per-SM basis.

\subsection{Predictor State}

\begin{table}
\centering
\scriptsize
\begin{tabular}{|c|c|}
\hline
\bf State & \bf Description \\ \hline
$Active\_Kernel\_Cycles$ & Cycles for which kernel is running on SM \\ \hline
$Done\_Blocks$ &  Number of thread blocks completed on SM \\ \hline
$Total\_Blocks$ &  Total number of thread blocks assigned to SM \\ \hline
$Resident\_Blocks$ &  Number of thread blocks resident \\ \hline
$Total\_Blocks\_Done$ &  Number of blocks completed so far \\ \hline
$Block\_Start[]$ & Cycle at which resident block started \\ \hline
$t$ & Duration of thread block \\ \hline
$Pred\_Cycles$ &  Total Runtime Cycles (Predicted) \\ \hline
$Reslice$ & Set to \code{true} when new slice has started \\ \hline
\end{tabular} 
\caption{State maintained per-kernel in each SM}
\label{tbl:smstate}
\end{table}

Table~\ref{tbl:smstate} details the state used by the Simple Slicing predictor that is maintained on each SM on a per-kernel basis.
State updates (except for the prediction itself) are independent of the predictor and occur on any of the following four events: launch of a kernel onto an SM ({\sc OnLaunch}), start of a thread block ({\sc OnBlockStart}), end of a thread block, ({\sc OnBlockEnd}) and finally end of a kernel ({\sc OnKernelEnd}).
Algorithms for state updates are detailed in Algorithm~\ref{alg:smstate}.

\begin{algorithm}
\footnotesize
\begin{algorithmic}[1]
\Function{OnLaunch}{$Kernel$}
\State $Resident\_Blocks \leftarrow Kernel.Residency$
\State $Total\_Blocks \leftarrow \lceil Kernel.Blocks / N_{SM} \rceil $
\State $Reslice \leftarrow {\rm true}$
\EndFunction

\Function{OnKernelEnd}{$Kernel$}
\For{All Running Kernels}
\State $Reslice \leftarrow {\rm true}$
\EndFor
\EndFunction

\Function{OnBlockStart}{$Kernel$, $blkindex$}
\State $Block\_Start[blkindex] \leftarrow clock()$
\EndFunction

\Function{OnBlockEnd}{$Kernel$, $blkindex$}
\State $Done\_Blocks++$
\If{$Reslice$}
\State $t \leftarrow clock() - Block\_Start[blkindex]$
\EndIf
\EndFunction

\end{algorithmic}
\caption{Functions to update per-kernel state on each SM. $Kernel.Residency$ is the maximum number of resident blocks for $Kernel$; $Kernel.Blocks$ is the total number of thread blocks; $N_{SM}$ is the number of SMs; $clock()$ is the current clock cycle; $blkindex$ is the block identifier on the SM (0--7).}
\label{alg:smstate}
\end{algorithm}

When a kernel is launched, we initialize all its per SM counters to zero. 
Then, $Resident\_Blocks$ is initialized to $R$, the maximum number of blocks that can reside at a time on an SM when running alone. %
Finally, we initialize $Total\_Blocks$ to the number of thread blocks we expect to execute on that SM.
With current schedulers, this is only an estimate. %
Current schedulers (e.g., Fermi) dynamically assign thread blocks to SMs. 
Depending on when each thread block terminates, the number of thread blocks executed per SM can vary.
$Total\_Blocks$ can, therefore, be less than or more than the actual number of blocks that execute on an SM.
We currently assume uniform distribution and hence set it to $\lceil Kernel.Blocks / N_{SM} \rceil$ where $N_{SM}$ is the number of SMs.

In {\sc OnBlockStart} and {\sc OnBlockEnd}, a number of book-keeping variables $Block\_Start$, $Done\_Blocks$ are updated.
$Active\_Kernel\_Cycles$ tracks the actual number of cycles the kernel has been running and is incremented on every cycle that it has a warp running on an SM.
$Pred\_Cycles$ contains the runtime prediction for the kernel and is calculated by the predictor using the following equation:
\begin{equation}
\begin{split}
Pred\_Cycles = & Active\_Kernel\_Cycles + \\  & \frac{(Total\_Blocks - Done\_Blocks)\times t}{Resident\_Blocks}
\end{split}
\label{eqn:ss}
\end{equation}

The prediction is calculated at the end of the handler of {\sc OnBlockEnd}, and uses the duration of the first thread block of a slice as the value for $t$.
The use of $Active\_Kernel\_Cycles$, which contains the actual runtime of a kernel so far, is to correct predictor drift.
$Reslice$ is set to \code{false} after every prediction.

\subsection{Predictor Accuracy}

We run traces of actual program runs through the predictor to evaluate the predictor's accuracy by comparing the predicted runtime to actual runtime.
Each trace of a program contains the start and end times for every thread block as well as the SM it ran on.
We group the traces as: (i)~single-gpu -- traces from runs of single applications on the GPU, (ii)~single-sim -- traces from runs of single applications on the simulator, (iii) mpmax -- traces from two-program workloads executed on the simulator using the JIT MPMax scheme described later in Section~\ref{sec:mpmax}.
The single-gpu and single-sim traces feature only a single slice whereas mpmax features at least two slices.
For mpmax, we measure accuracy only for the last slice.

\begin{figure}
\centering
\includegraphics[scale=0.45]{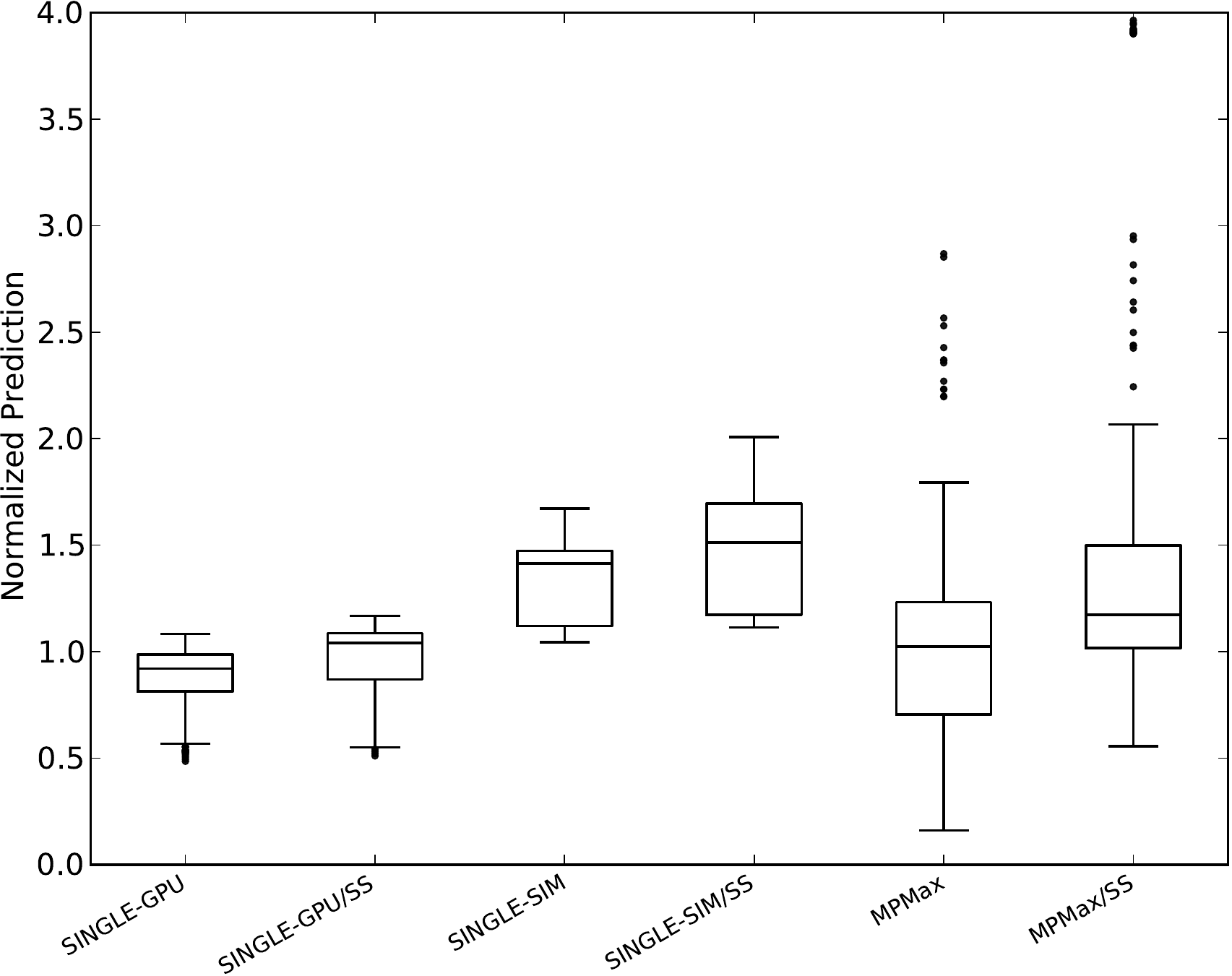}
\caption[Accuracy of the Simple Slicing (SS) Predictor]{Accuracy of the Simple Slicing (SS) Predictor. Runtime Predictions normalized to actual runtime.}
\label{fig:pred_accu_results}
\end{figure}

We evaluate Equation~\ref{eqn:ss} in a slice-aware mode (``/SS'') as well as a slice-unaware mode, where the prediction is only made once, at the beginning of the kernel.
Figure~\ref{fig:pred_accu_results} presents the the results.
For all the groups, the Simple Slicing predictor is accurate to within 2x of the actual runtime for the majority of programs.
For single-gpu, the predictions are between 0.48x to 1.08x of actual runtime.
Since Equation~\ref{eqn:ss} is not step function, single-sim predictions are less accurate than those for the hardware.
For mpmax, the simple slicing predictor corrects the underestimates made by the slice-unaware predictor, and the majority of its predictions are between 0.5x and 2x of runtime.
We emphasize that these errors do not limit our scheduling policies as our overall evaluation will show.

\section{Thread Block Scheduling}
\label{sec:tbs}

The four scheduling policies that we evaluate in this work consist of two of our policies, SRTF and SRTF/Adaptive, both of which use estimates of runtime  provided by the Simple Slicing predictor to guide their scheduling decisions and two other policies FIFO and MPMax which do not use runtime estimates.
Other than FIFO, all the policies are concurrent-kernel aware.

\subsection{Runtime Aware Policies}
 
\subsubsection{Shortest Remaining Time First (SRTF)}
\label{sec:srtf}
\begin{figure}
\centering
\includegraphics[scale=1]{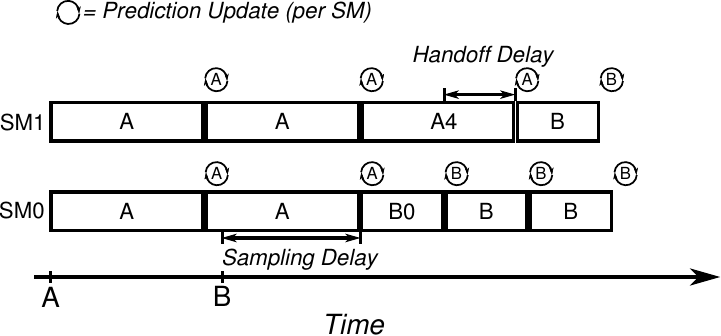}
\caption[Timeline of TBS Actions]{Timeline of TBS Behaviour for Kernels A and B under SRTF}
\label{fig:srtf}
\end{figure}

Figure~\ref{fig:srtf} depicts an example timeline of our TBS executing our SRTF policy.
Initially, kernel $A$ begins execution. 
As each of its thread blocks completes, we update the prediction for $A$'s runtime on each SM.
After concurrent kernel $B$ launches, we need a prediction to decide if it must replace $A$.

To do this with minimum disruption and delay, we sample $B$ on a single SM (here, SM0).
Essentially, after a {\it sampling delay\/}, when $B$ must wait for $A$'s thread blocks to complete, $B$ begins execution (shown as B0).
Meanwhile, $A$ continues executing on the other SMs.
Since the other SMs can execute $A$'s thread blocks intended for SM0, sampling $B$ is minimally disruptive.

Once a sufficient number of $B$'s blocks finish, a sample prediction can be made on SM0.
In this example, $B$'s sample prediction indicates that it will finish faster than $A$, so SM0 switches to $B$.
The sample prediction is copied to the other SMs as an initial prediction.
These SMs, after a {\it hand-off delay\/} during which $A$'s blocks are still executing, also switch to executing $B$'s thread blocks.

As $B$'s execution continues, each SM continues to refine $B$'s prediction.
If executing $B$'s thread blocks is no longer correct (say, if $B$'s later thread blocks do more work than $B$'s initial thread blocks), the SMs will switch back to $A$.

Our SRTF implementation only samples a single kernel at a time.
Further, when not sampling, only a single kernel executes on the GPU.
When multiple concurrent kernels are queued up, each is sampled in FIFO order, with the goal being to execute the shortest kernel first.

\subsubsection{SRTF/Adaptive}
\label{sec:adaptive}
\begin{figure}
\centering
\includegraphics[width=0.45\textwidth]{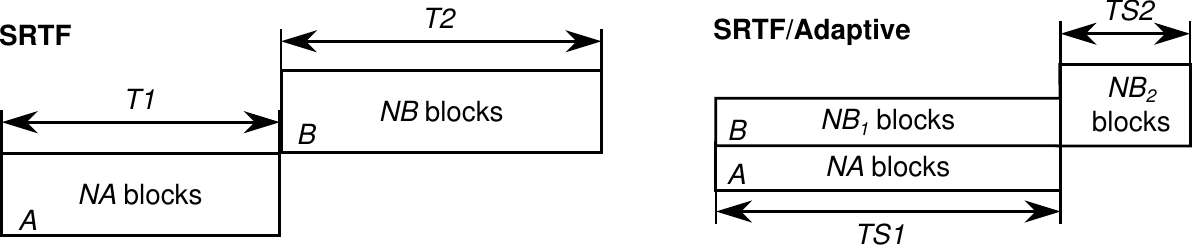}
\caption[SRTF versus SRTF/Adaptive]{SRTF versus SRTF/Adaptive. Note that $NB_1 + NB_2 = NB$. Illustrative, not to scale.}
\label{fig:adaptive}
\end{figure}
The SRTF scheduling policy does not share resources among concurrently executing kernels.
While this leads to good performance, it can be unfair.
In the example of SRTF scheduling depicted in Figure~\ref{fig:adaptive}, kernel $B$ is delayed by $T1$, the time taken by kernel $A$, leading to a slowdown of $(T1 + T2) / T2$.
In the extreme case, if $T1$ is only slightly smaller than $T2$, then $B$ experiences nearly a 2x slowdown.
By sharing resources between the two kernels, as in the SRTF/Adaptive part of the same figure, it is possible to ensure equitable progress for both kernels.
The slowdown in shared execution for $A$ and $B$ is then $TS1/T1$ and $(TS1+TS2)/T2$ respectively.

Our proposal for resource sharing, SRTF/Adaptive, shares GPU resources among concurrently executing kernels, but only if it detects that the current SRTF schedule is unfair.
To do this, it calculates the slowdown for running kernels in non-sharing mode as above.
If the difference between the smallest slowdown and the largest slowdown exceeds a threshold (we use 0.5), SRTF/Adaptive switches to sharing mode in which the maximum residency of kernel is limited with the rest turned over to co-running kernels.

Once in sharing mode, SRTF/Adaptive continues to monitor slowdowns for each program.
Calculation of shared runtimes (as in Figure~\ref{fig:adaptive}) is slightly more involved.
The value of $TS1$ is simply the runtime for $A$.
Calculating $TS2$ requires knowing $NB_1$ and $NB_2$, the number of blocks of $B$ that execute in shared and exclusive mode respectively.
$NB_1$ is obtained by solving for $N$ using Equation~\ref{eqn:total} with $R$ equal to the shared residency and $T = TS1$ (i.e. $N = TR/t$).
The value of $NB_2$ is then $N - NB_1$.
This is an iterative procedure over all running kernels and is hence bounded by the limited number of concurrent kernels (8 on the NVIDIA Fermi). %
The exclusive runtime (i.e. $T1$ or $T2$) is the prediction from the exclusive part of a run.
If there was no exclusive part, current predictions of runtime are used instead.

Determining the split of resources between the co-running kernels is harder.
We evaluated an implementation that initially distributed resources equally among all running kernels and then redistributed resources as necessary.
However, it was too slow to achieve both fair execution and good performance on our workloads since redistribution could only happen at thread block completion.
Therefore, we chose a fixed residency limit of 3, one less than half of the resources, for the fastest kernel.
Static partitioning is not optimal for some workloads, but it works well on average.

\subsection{Runtime Unaware Policies}

\subsubsection{FIFO (Baseline)}
\label{sec:fifo}
The FIFO Thread Block Scheduler is based on the NVIDIA Fermi Thread Block Scheduler.
It schedules thread blocks from running kernels in the order of their arrival.
Only when all the thread blocks of a kernel have been dispatched to the streaming multiprocessors for execution are blocks from the next kernel scheduled.

\subsubsection{Just-in-Time MPMax}
\label{sec:mpmax}
Just-in-time MPMax is a resource-allocation policy, not a scheduling policy {\it per se}.
It is based on the best-performing MPMax resource-allocation policy~\cite{pai2013}.
In this policy, each executing kernel sets aside resources for a hypothetical ``MPMax'' kernel, a composite constructed from co-running kernels.
For example, under this policy, when two kernels $A$ and $B$ execute together, $A$ will set aside resources for one thread block of $B$ per SM and vice versa.
Our Just-in-time adaptation improves on the original MPMax: i) the resources set aside by each kernel are calculated on the basis of the kernels actually running on the GPU at that instant and ii) when concurrent kernel execution ceases, kernels reoccupy the resources they had set aside.
When scheduling, thread blocks are issued from a kernel until its MPMax limit is reached, after which the next kernel in FIFO order gets to issue thread blocks.

\section{Evaluation}
\label{sec:tbseval}
We evaluate the execution of concurrent kernels under the four TBS policies discussed in Section~\ref{sec:tbs}.
Since existing benchmarks lack concurrent kernels, our evaluation uses 2-program workloads.
The primary metrics reported are system throughput (STP), average normalized turnaround time (ANTT)~\citep{eyerman2008} and the StrictF metric for fairness~\citep{vandierendonck2011}.
StrictF is defined as the ratio of minimum slowdown to maximum slowdown, with a value of 1 indicating high fairness.

\subsection{Experimental Setup}

\subsubsection{Benchmarks}

\begin{table}
\footnotesize
\centering

\begin{tabular}{|c|c|c|c|c|}
\hline
\bf Benchmark       & \bf Kernel & \bf R      & \bf TPB    & \bf Blocks\\
\hline
AES-d   &aesDecrypt128  &6      &256    &1429\\
\hline
AES-e   &aesEncrypt128  &6      &256    &1429\\
\hline
ImageDenoising-nlm2     &NLM2   &8      &64     &4096\\
\hline
JPEG-d  &CUDA...IDCT        &8      &64     &512\\
\hline
JPEG-e  &CUDA..DCT &8      &64     &512\\
\hline
RayTracing      &render\footnotemark &5     &128    &2048\\
\hline
SAD     &mb\_sad\_calc  &8      &61     &1584\\
\hline
SHA1    &sha1\_kernel\_direct   &8      &64     &1539\\
\hline
\end{tabular}
\caption[Grid Characteristics of ERCBench Kernels]{Grid Characteristics of ERCBench Kernels. TPB = Threads Per Block, R = Maximum Residency}
\label{tbl:erc1}
\end{table}

\begin{table}
\scriptsize
\centering
\begin{tabular}{|c|c|c|c|c|}
\hline
\multirow{2}{*}{\bf Benchmark} & \multirow{2}{*}{\bf Kernel}     &\multicolumn{3}{|c|}{\bf Simulator (cycles)}\\
\cline{3-5}
&       &\bf Runtime       & \bf Mean $t$  &\bf \%RSD\\
\hline
AES-d   &aesDecrypt128     &234154 &14529        &12.52\\
\hline
AES-e   &aesEncrypt128      &226335 &14031        &12.1\\
\hline
ImageDenoising-nlm2 & NLM2        &692686 &19873        &2.87\\
\hline
JPEG-d  &CUDA..IDCT   &24853  &5238        &29.58\\
\hline
JPEG-e  &CUDA...DCT    &25383  &5367        &32.95\\
\hline
RayTracing & render     &416563 &15167        &65.71\\
\hline
SAD & mb\_sad\_calc   &441297 &32332        &6.57\\
\hline
SHA1 & sha1\_kernel...     &22224223        &1708531        &7.98\\
\hline
\end{tabular}
\caption[Runtimes for ERCBench Kernels on the simulator]{Runtimes for ERCBench Kernels on the simulator. \%RSD = $100 \times {\rm Std. Dev}(t)/{\rm Mean}\,t$}
\label{tbl:erc2}
\end{table}

Our 2-program workloads are composed of 8 kernels from 8 ERCBench~\citep{chang2010} benchmarks.\footnotetext{The instrumented version of \code{render} used in Figure~\ref{fig:tbtimes_residency} uses one fewer register allowing it to have six resident blocks. This is an artefact of the CUDA compiler.}
The DVC, RSA and Bitonic benchmarks are not used in our evaluation because the first two do not run on our simulator and BitonicSort runs using only one thread block.
Tables~\ref{tbl:erc1} and~\ref{tbl:erc2} highlight the grid characteristics and runtimes for ERCBench kernels and will be used to interpret our results in the following sections.

\subsubsection{Simulator}
\label{sec:simulator}
\begin{table}
\scriptsize
\centering
\begin{tabular}{|l|p{5cm}|}
\hline
Number of SMs & 15 \\
\hline
Resources per SM & 1536 threads, 32768 registers, 48KB shared memory, Maximum 8 Thread Blocks, Maximum 48 warps  \\
\hline
Threads per warp & 32 \\
\hline
Warp scheduler & Loose Round Robin \\
\hline
\end{tabular}
\caption{Simulator Configuration}
\label{tbl:simconfig}
\end{table}
We modify the GPGPU-Sim simulator~(3.2.0)~\citep{gpgpusim} extending the simulator to execute multiple kernels concurrently (the released version only runs a single kernel on an SM at a time) but leave the actual cycle-accurate simulator for each thread unchanged.
We also add a functional thread block scheduler and predictors whose behaviour is as described in Sections~\ref{sec:onlinepred} and \ref{sec:tbs}.
The simulated GPU configuration, listed in Table~\ref{tbl:simconfig}, is the GTX 480 configuration supplied with GPGPU-Sim.

\subsubsection{Methodology}

Multiprogrammed workloads cannot be run directly on the GPU or on the simulator.
Therefore, to achieve concurrent execution of kernels, we use the techniques described in~\cite{pai2013} to construct multithreaded workloads with each thread executing the individual CUDA programs.
As GPGPU-sim performs cycle-accurate simulation only for GPU kernels with memory transfers only functionally emulated~\citep{lustig2013} and CPU code running at full speed, we only record timings for the simulated kernels even though we run the entire workload to completion.

We use all possible 28 2-program workloads from the ERCBench suite.
Since the order of kernel arrivals affects a scheduler significantly, we simulate and present results for both orders of arrival making for a total of 56 2-program workloads.
Our primary results evaluate kernel arrivals that are staggered by upto 100 cycles, thus the two kernels start nearly together.
We also present results for different arrival offsets where the second kernel arrives after 25\% and 50\%  of the first kernel has finished executing.

\subsection{Results}

\subsubsection{Overall Results}

\begin{table}
\scriptsize
\centering
\begin{tabular}{|c|c|c|c|}
\hline
\bf Scheduler  &  \bf STP  &  \bf ANTT & \bf Fairness\\ 
\hline
\policy{FIFO} & 1.35 & 3.66 & 0.19 \\ \hline
\policy{MPMax} & 1.37 & 2.15 & 0.36 \\ \hline
\policy{SRTF} & 1.59 & 1.63 & 0.52 \\ \hline
\policy{SRTF/Adaptive} & 1.51 & 1.64 & 0.56 \\ \hline
\policy{SJF} & 1.82 & 1.13 & 0.80 \\ \hline
\end{tabular}
\caption[Geomean STP, ANTT and Fairness for various scheduling policies]{Geomean STP, ANTT and Fairness for various scheduling policies. Note that ANTT is a lower-is-better metric. }
\label{tbl:tbsstpantt}
\end{table}

Table~\ref{tbl:tbsstpantt} summarizes the results of our evaluation.
SJF shows the best STP, ANTT and fairness values, but is unrealizable.
Next to \policy{SJF}, the \policy{SRTF} policy has the best STP and ANTT among all scheduling policies.
It also has the second best fairness value among the realizable policies that we evaluated.
Compared to our baseline \policy{FIFO}, \policy{SRTF} improves STP by 1.18x, ANTT by 2.25x and Fairness by 2.74x.
\policy{SRTF} also outperforms \policy{MPMax} by 1.16x (STP) and 1.3x (ANTT).
The \policy{Adaptive} policy is the fairest among all the realizable policies studied, with a 2.95x fairness improvement over \policy{FIFO}.
It is also the second-best policy with its STP being 1.12x better and ANTT being 2.23x better than baseline \policy{FIFO}.
However, since \policy{Adaptive} achieves fairness by sharing resources between concurrently executing kernels, its STP is lower by 5\% than that of \policy{SRTF}.
\policy{FIFO} is the least fair policy.

\subsubsection{System Throughput}
\label{sec:detailedstp}
\begin{figure}
\centering
\includegraphics[width=0.4\textwidth]{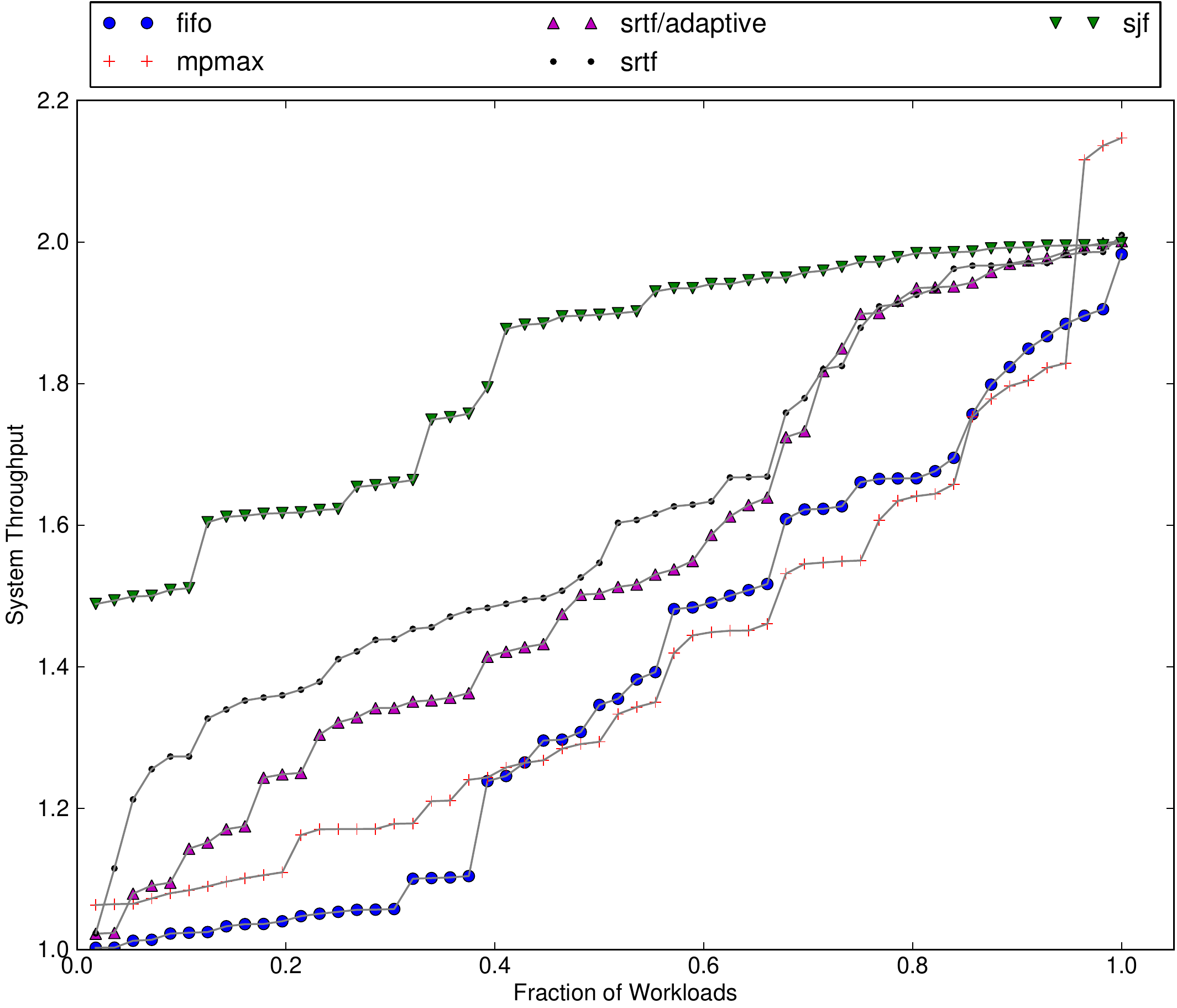}
\caption{System Throughput (STP) for various scheduling policies}
\label{fig:tbsallstp}
\end{figure}

Figure~\ref{fig:tbsallstp} plots the system throughput for all 56 workloads for all policies.
\policy{SRTF} outperforms other non-SJF schedulers in nearly all of the workloads.
However, as Table~\ref{tbl:tbsstpantt} shows, there is a gap of 12.64\% between \policy{SRTF} and \policy{SJF}.
Unlike \policy{SJF} which schedules kernels even before they run, \policy{SRTF} must {\it learn\/} the runtimes of concurrently executing kernels to determine which is the shorter kernel.
In our implementation, this is done through a sample execution as described in Section~\ref{sec:tbs}.
Since running thread blocks cannot be preempted, sample execution must wait until resources are available for sampling.
This leads to two possible scenarios.
In the first scenario, the kernel currently executing has the shortest runtime, and hence the sampling procedure disrupts its execution when compared to \policy{SJF}.
In the second scenario, the latest kernel to arrive has the shorter runtime and so the time it waits for sampling to begin delays its execution as compared to \policy{SJF}.

The effect of sampling on performance is largely determined by the relative thread block durations of each concurrently executing kernel.
Consider the \name{RayTracing+JPEG-d} pair in which \name{JPEG-d} is launched as the second kernel.
\name{JPEG-d}'s kernel is very small, with average thread block duration of 5238 cycles (Table~\ref{tbl:erc2}).
\name{RayTracing}'s thread blocks take on average 15168 cycles to execute.
Arriving second, \name{JPEG-d}'s worst-case sampling delay is therefore on average 15168 cycles.
Once sampling is done, the worst-case handoff delay is also on average 15168 cycles on the SMs that were not participating in the sampling and which would have continued executing thread blocks from \name{RayTracing}.
So, for a kernel that takes a total of about 25000 cycles when running alone, in this example \name{JPEG-d} has already slowed down by 2x. 
This is still better than the 17.76x slowdown under FIFO.

To quantify the effects of sampling on performance, we conducted an experiment where we omitted the sampling phase.
In this {\it zero-sampling\/} variant of SRTF, we provided the runtimes to the \policy{SRTF} scheduler directly as in \policy{SJF}.
For our workloads, this improved STP by 3\% to 1.64 and ANTT by 22\% to 1.33.
The remaining performance gap is therefore only due to the hand-off delay.
Therefore the inability to preempt running thread blocks is thus the major performance limiter for scheduling on the GPU.
Since the zero sampling experiment also provided accurate runtimes to the \policy{SRTF}, the results also show that \policy{SRTF} is very tolerant of errors in the simple slicing predictor.

On average, \policy{MPMax} and \policy{FIFO} have almost the same throughput.
The detailed results show that \policy{MPMax} outperforms \policy{FIFO} for slightly more than 50\% of the workloads.
This is because of our experimental setup which evaluates all possible 2-program workloads.
For half of the workloads, \policy{FIFO} schedules just as \policy{SJF} would.

\policy{MPMax} performs better than SJF for three workloads.
In all three workloads, we find that SHA1 executes second and finishes approximately 21\% to 37\% faster than when running alone.
Our experiments on real hardware with {\it non-JIT\/} \policy{MPMax}~\cite{pai2013} failed to exhibit this speedup, though we have observed such speedups in shared mode on hardware.

\subsubsection{Average Normalized Turnaround Time}
\label{sec:detailedantt}
\begin{figure}
\centering
\includegraphics[width=0.4\textwidth]{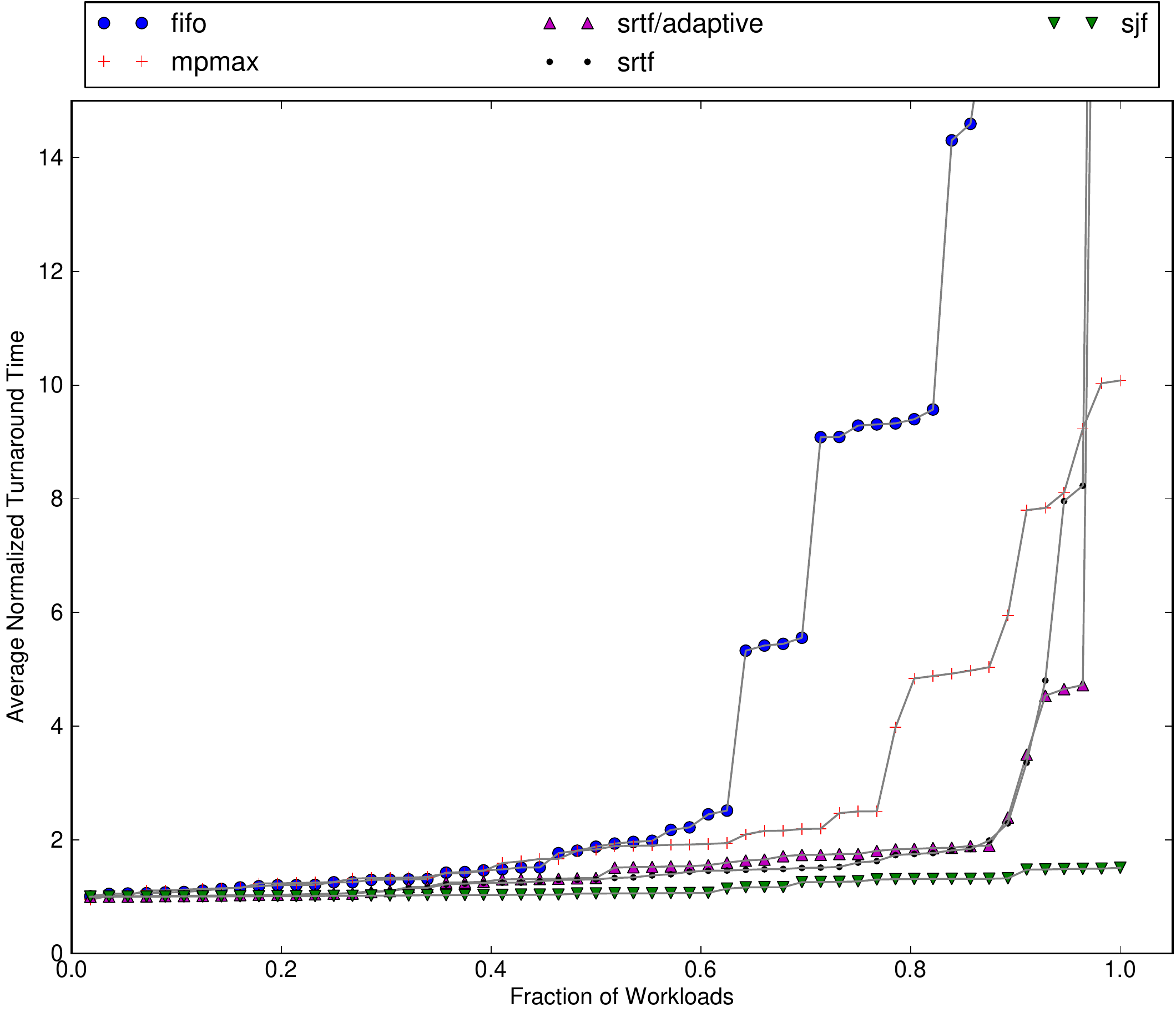}
\caption[Average Normalized Turnaround Time (ANTT) for various scheduling policies]{Average Normalized Turnaround Time (ANTT) for various scheduling policies. ANTT is lower-is-better metric.}
\label{fig:tbsallantt}
\end{figure}

Figure~\ref{fig:tbsallantt} shows that the ANTT values are nearly indistinguishable for about 35\% of the benchmarks, \policy{SRTF} and \policy{Adaptive} are the only realizable policies to have the lowest ANTT values for all but two of the workloads.
At 30.95 and 37.77, the worst ANTT values for \policy{SRTF/Adaptive} and \policy{SRTF} (not shown in the figure) are well below the worst ANTT value (\policy{FIFO} with 425.45) but are still higher than \policy{MPMax} whose worst ANTT value is 10.08.
These maximum values are for \name{SHA1+JPEG-d} (the next highest value is for \name{SHA1+JPEG-e}) and are the result of \name{JPEG-e} having to endure a hand-off delay of 1.7M cycles.
Since \policy{MPMax} reserves runtime resources on all SMs for concurrently executing kernels as soon as they launch, \name{JPEG-e} does not experience hand-off delay.

\subsubsection{Fairness}

\begin{figure}
\centering
\includegraphics[width=0.4\textwidth]{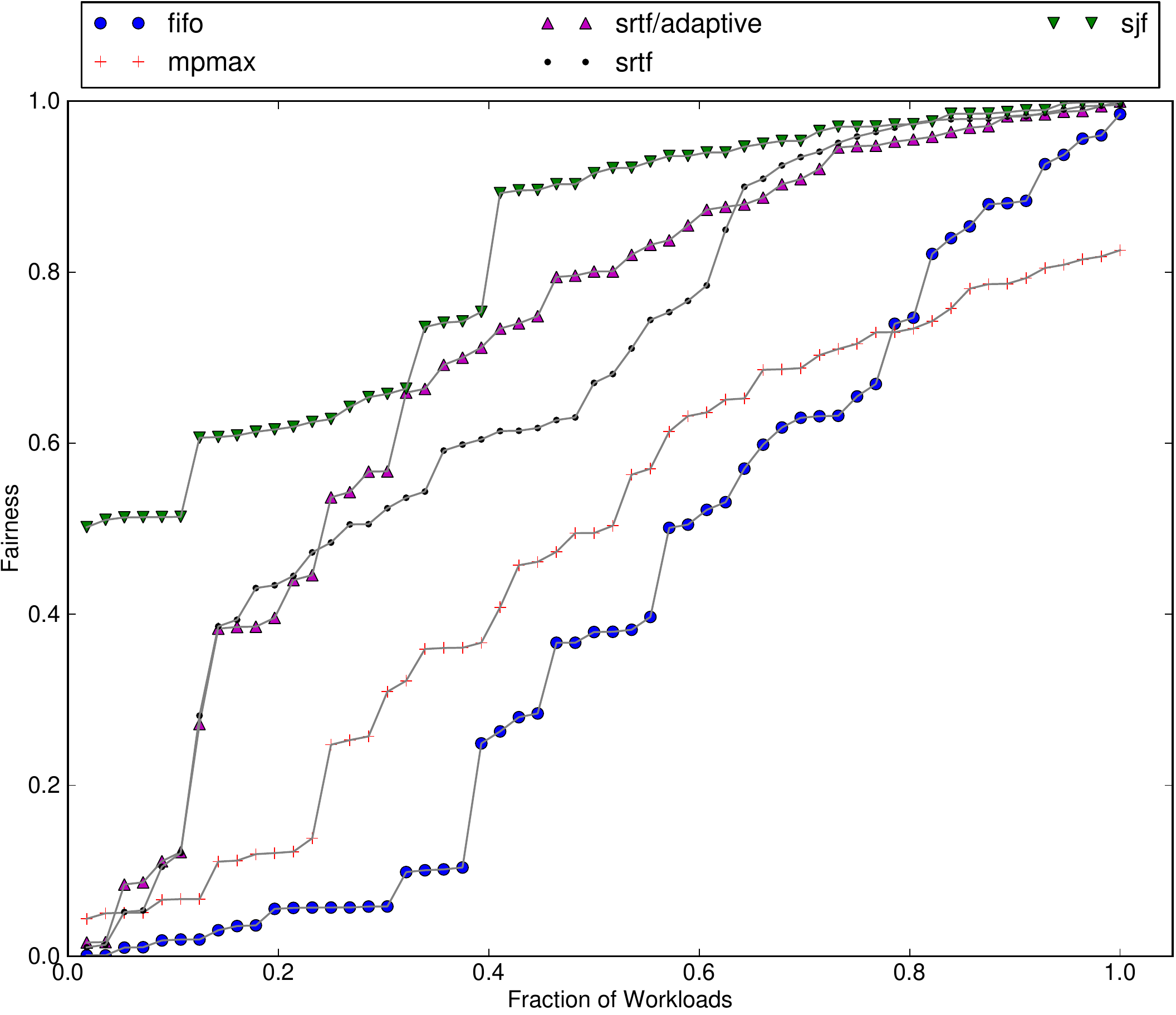}
\caption{Fairness (StrictF) for various scheduling policies.}
\label{fig:tbsallfairness}
\end{figure}

Figure~\ref{fig:tbsallfairness} plots StrictF, the fairness metric we use, for all of the workloads.
\policy{SRTF/Adaptive}, our fairness-oriented policy, executes 35 of the 56 workloads in sharing mode.
In 34 of the 56 workloads, it achieves higher fairness than any the other realizable policies.
System throughput under \policy{SRTF/Adaptive} is within 5\% of \policy{SRTF}, while ANTT is nearly the same.

While \policy{MPMax} achieves higher fairness than \policy{FIFO} for 80\% of the workloads, for 20\% of the workloads, however, sharing leads to a loss in performance that is relatively greater for the smaller component of the workload as compared to running with full resources.

\subsubsection{Sensitivity to Arrival Time}

To evaluate sensitivity to arrival time, we simulate workloads where the second kernel arrives after the first kernel has executed a fixed number of cycles.
Table~\ref{tbl:offsetstpantt} presents the results when the second kernel arrives at 25\% and 50\% of the runtime of the first kernel.
\policy{SRTF} continues to perform well across all metrics for both 25\% and 50\% arrival offsets.
From 25\% to 50\%, however, the gaps between the different policies shrink, a consequence of the fact that as the kernels start farther apart in time, there is less opportunity for the scheduler to perform.

\begin{table}
\scriptsize
\centering
\begin{tabular}{|c|c|c|c|c|c|c|}
\hline
\multirow{2}{*}{\bf Scheduler} & \multicolumn{3}{|c|}{\bf 25\%} & \multicolumn{3}{|c|}{\bf 50\%} \\
\cline{2-7} 
 &  \bf STP  &  \bf ANTT & \bf Fair. &  \bf STP  &  \bf ANTT & \bf Fair. \\ 
\hline
\policy{FIFO} & 1.44 & 2.74 & 0.27 & 1.48 & 2.36 & 0.32 \\ \hline
\policy{MPMax} & 1.45 & 2.05 & 0.38 & 1.49 & 1.93 & 0.40 \\ \hline
\policy{SRTF} & 1.62 & 1.60 & 0.53 & 1.63 & 1.56 & 0.55 \\ \hline
\policy{SRTF/Adaptive} & 1.56 & 1.65 & 0.56 & 1.59 & 1.58 & 0.59 \\ \hline
\end{tabular}
\caption[Geomean STP, ANTT and Fairness for various scheduling policies when second kernel arrives at 25\% and 50\% of first kernel's runtime]{Geomean STP, ANTT and Fairness for various scheduling policies when second kernel arrives at 25\% and 50\% of first kernel's runtime. ANTT is a lower-is-better metric. }
\label{tbl:offsetstpantt}
\end{table}

\section{Related Work}
\label{sec:tbsrelated}

\citet{minseoklee2014} recently proposed a thread block scheduler that throttles the number of thread blocks executing on an SM based on performance.
The freed resources are allocated to concurrently running kernels.  
Their scheme does not preempt the running kernel, nor change the order of running kernels.

\citet{nath2013} proposed $T^ABS$, which interleaves thread blocks from concurrent kernels to manage thermal emergencies on GPUs.
Based on online profiling of thermal characteristics, their thread block scheduler (in conjunction with an OS-level scheduler) takes away resources from ``hot'' kernels in the event of thermal emergencies, redistributing to ``cold'' kernels.
Their work illustrates a complementary goal achieved by the TBS.

TimeGraph~\citep{kato2011} schedules OpenGL programs at the OS-level.
It enforces policies by limiting access to the GPU at the device driver level using OS-level priorities.
TimeGraph is not preemptive and does not support concurrent kernels.

Runtime prediction of GPU kernel execution time for scheduling across heterogeneous CPU/GPU systems has been explored in several works~\citep{belviranli2013,jia2012,gregg2011,luk2009,jimenez2009,diamos2008}.
Although we do not explore scheduling a kernel across CPU and GPU, our online predictor does not require a historical database and is also aware of concurrent kernels.

\section{Conclusion}
\label{sec:tbsconclusion}

We presented a novel online runtime predictor for GPU kernels that exploited the structure of kernel grids to obtain runtime predictions by observing thread block durations.
We used it to build a thread block scheduler with runtime aware policies, SRTF and SRTF/Adaptive, which were found superior to the other realizable policies in terms of system throughput, turnaround time and fairness.
Compared to FIFO, our SRTF policy improved STP by 1.18x and ANTT by 2.25x.
SRTF also outperformed MPMax, a state-of-the-art resource allocation policy, with improvements of 1.16x in STP and 1.3x in ANTT.
Our SRTF/Adaptive policy achieved the highest fairness among all the realizable policies, 2.95x better than FIFO.
Finally, SRTF bridged 49\% of the gap between FIFO and SJF, approaching to within 12.64\% of SJF's throughput.

\newcommand{\BIBdecl}{\setlength{\itemsep}{0pt}\footnotesize}
\bstctlcite{bstctl:etal, bstctl:nodash, bstctl:simpurl}
\bibliographystyle{IEEEtranSN}
\bibliography{paper}

\section{Supplementary results on the NVIDIA Kepler}
\label{sec:kepler}
We repeated the experiments of Section~\ref{sec:structural} on a Kepler-based NVIDIA Kepler K20Xm.
The host machine is an octocore Intel Xeon E5-2609 (2.4GHz) and runs the 64-bit version of Debian Linux 7.1 with CUDA driver~319.32 and CUDA runtime~4.2.
Since there is no cycle-accurate simulator that models the Kepler GPU, we continue to use the Fermi in our evaluation in our paper.

\begin{figure}
\centering
\includegraphics[scale=0.45]{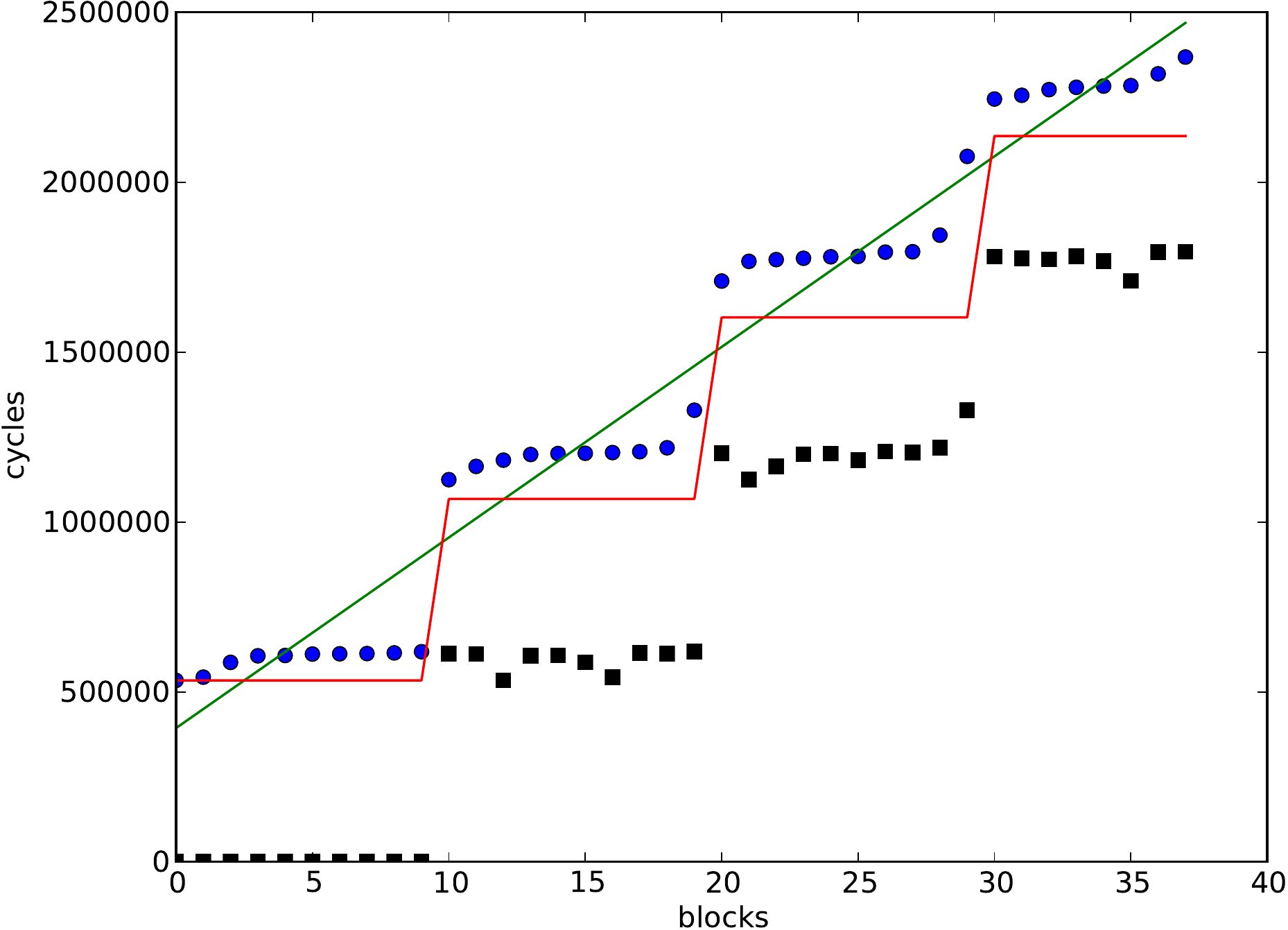}
\caption[Execution of SGEMM's thread blocks from one SM]{Execution of SGEMM's thread blocks on one SM. Blocks are ordered by finishing time. Black squares represent start times of each thread block, dark blue circles denote ending time. Green line is linear fit to all the end timings. Red line is prediction from equation~\ref{eqn:total}, with $t$ being the duration of the first block to finish. (Kepler equivalent of Figure~\ref{fig:sgemm})}
\label{fig:kepler:sgemm}
\end{figure}

Figure~\ref{fig:kepler:sgemm} shows that the staircase model continues to hold on the Kepler for SGEMM.
In fact, on the Kepler, we cannot find instances of staggered execution of SGEMM on other SMs as in Figure~\ref{fig:sgemm_inacc}.
However, other benchmarks do exhibit staggered execution.

\begin{figure}
\centering
\includegraphics[scale=0.45]{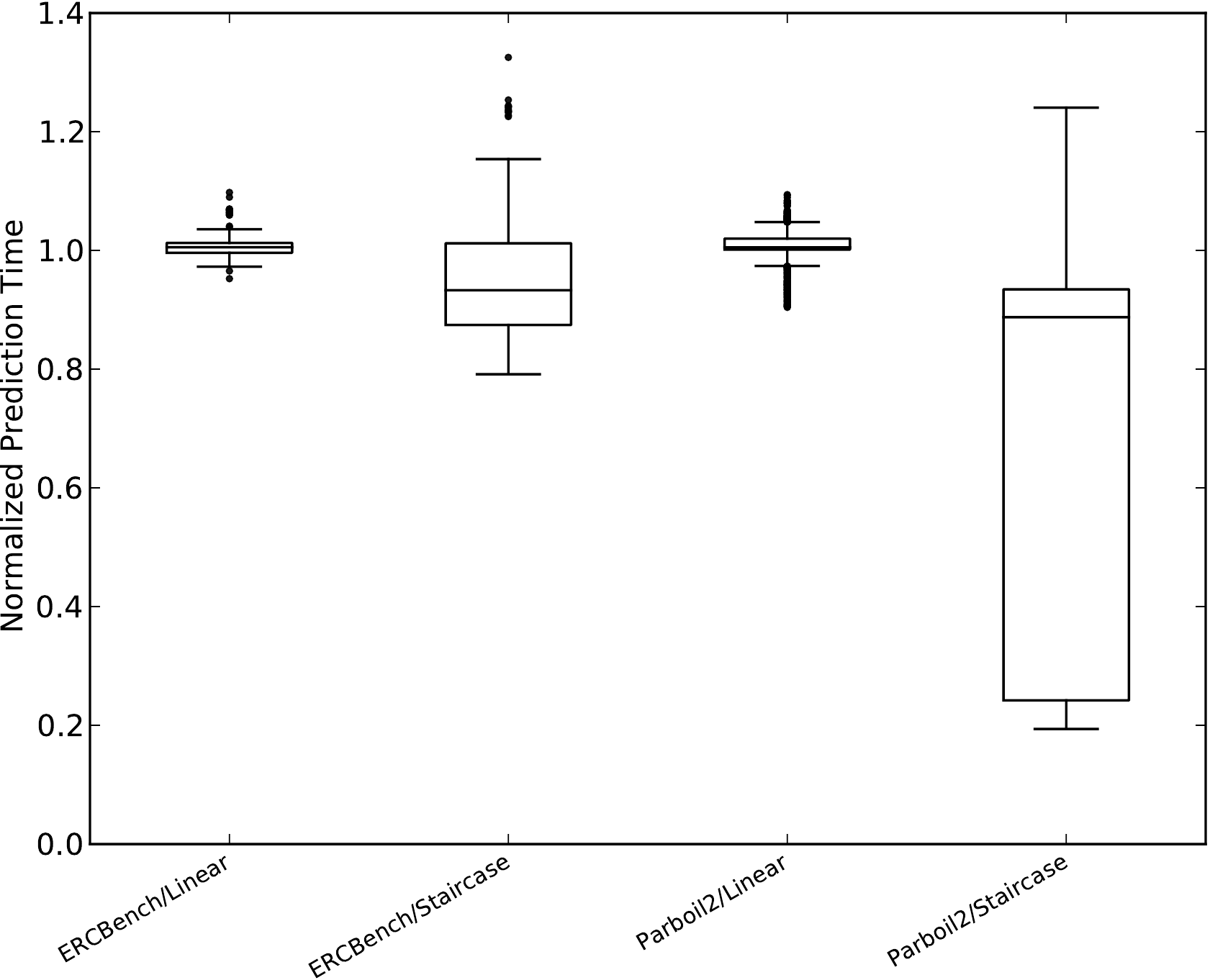}
\caption[Boxplots of Predictions from Linear Regression and Staircase Models]{Boxplots of Predictions from Linear Regression and Staircase Models for ERCBench and Parboil2 benchmarks normalized to actual runtime. (Kepler equivalent of Figure~\ref{fig:linear_model})}
\label{fig:kepler:linear_model}
\end{figure}

Figure~\ref{fig:kepler:linear_model} uses 4550 (the 28 extra predictions are from two additional iterations of LBM) predictions for Parboil2 kernels and 112 predictions for the kernels in ERCBench.
Linear regression results in normalized predictions between 0.95x to 1.09x of actual runtime for ERCBench and 0.9x to 1.09x for Parboil2.
Predictions from Equation~\ref{eqn:total} normalized to actual runtime lie between 0.79x to 1.33x for ERCBench and 0.19x to 1.24x for Parboil2.
The LBM benchmark in Parboil2 exhibits staggered execution on the Kepler, unlike the Fermi, and hence all of its 1400 predictions are underestimates.

\begin{figure}
\centering
\includegraphics[scale=0.45]{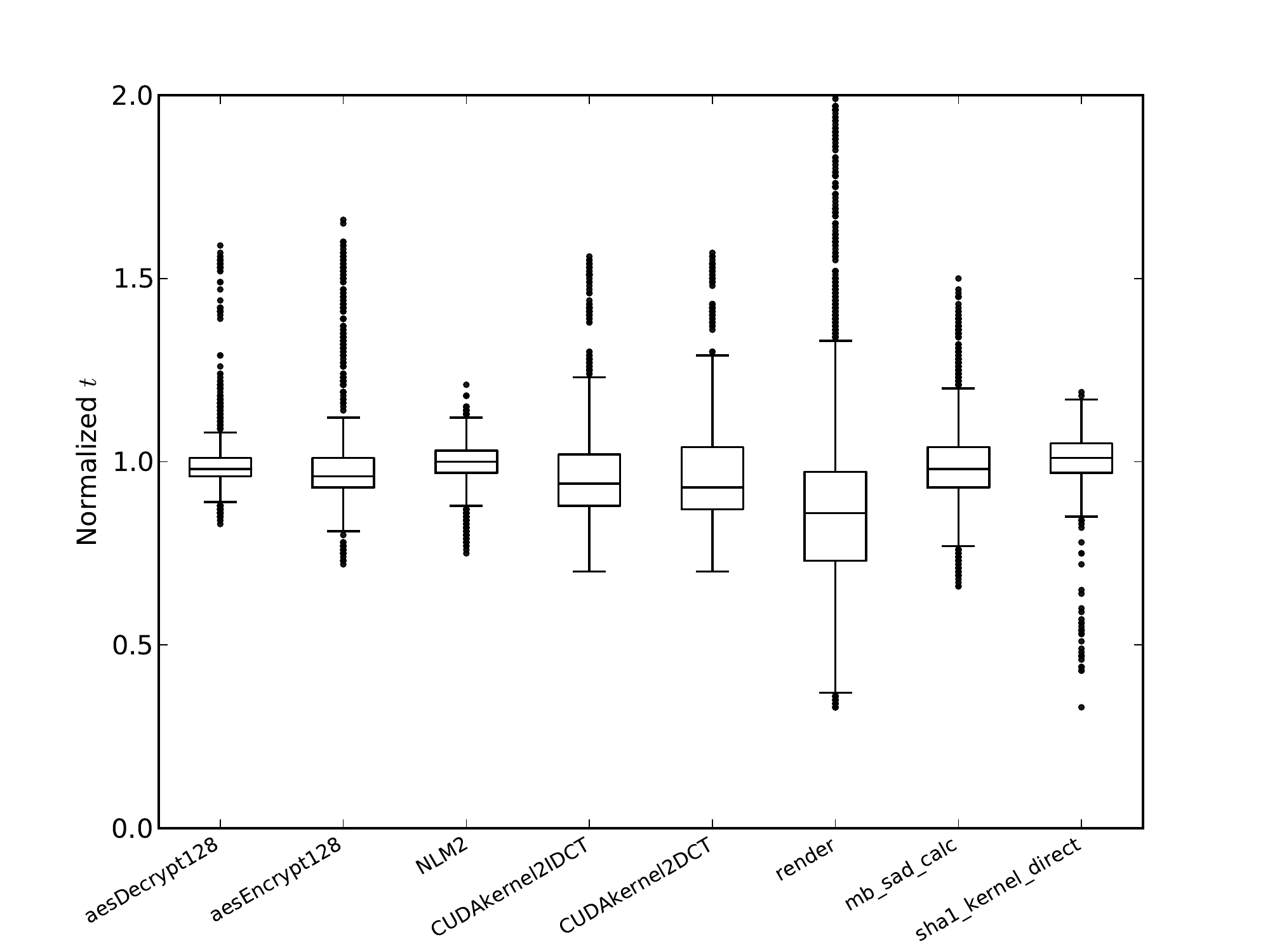}
\caption[Boxplots of normalized thread block durations]{Boxplots of thread block durations ($t$) normalized to their average for a kernel. \code{render}'s maximum value is 3.66. (Kepler equivalent of Figure~\ref{fig:blocktimespread})}
\label{fig:kepler:blocktimespread}
\end{figure}

\begin{figure}
\centering
\includegraphics[scale=0.38]{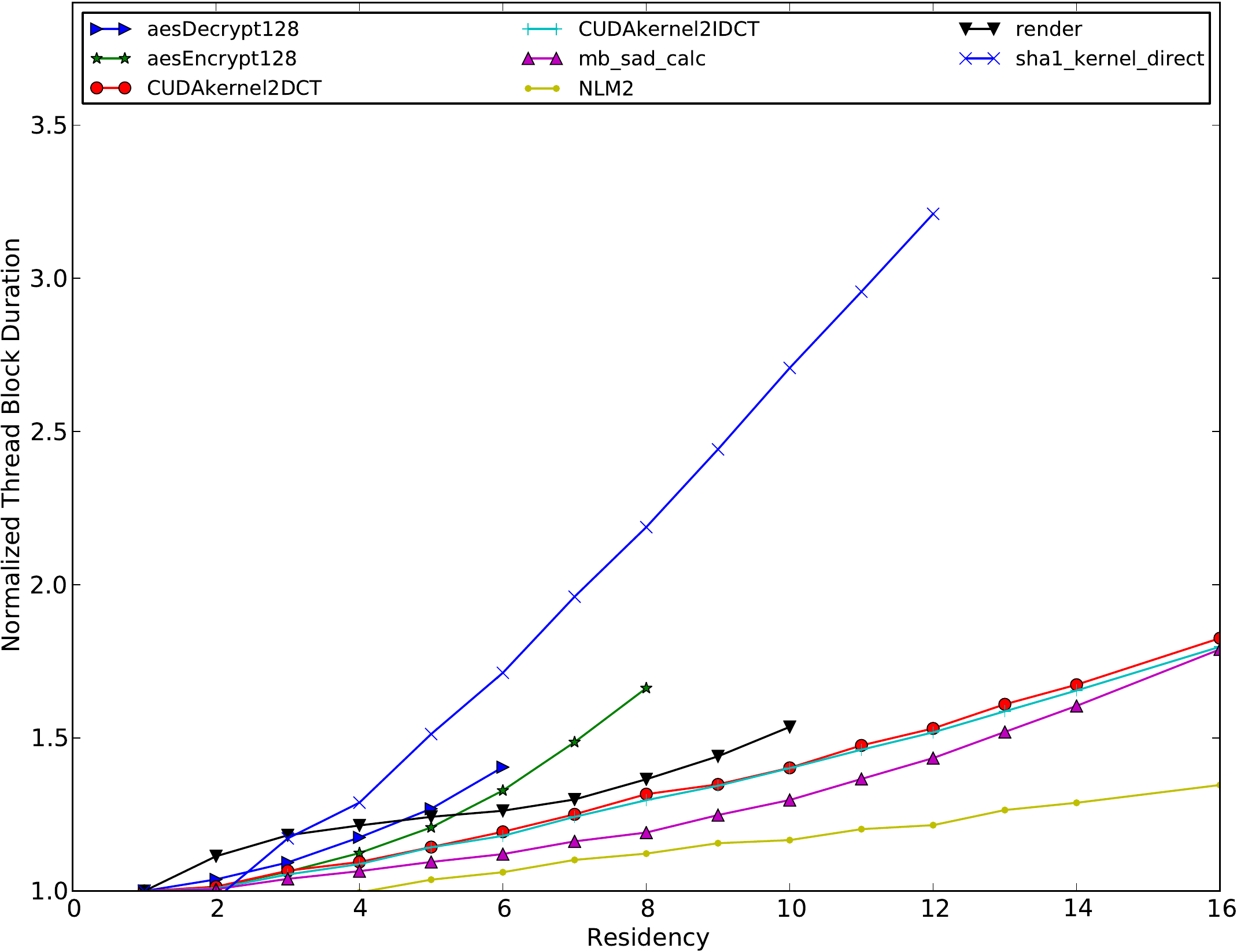}
\caption[Average thread block duration at various residencies]{Average thread block duration at various residencies normalized to average thread block duration at residency~1. Residencies are: AES-e (8), AES-d (6), RayTracing (10), SHA1 (12) and all other kernels have the maximum residency of 16 thread blocks. (Kepler equivalent of Figure~\ref{fig:tbtimes_residency})}
\label{fig:kepler:tbtimes_residency}
\end{figure}

\begin{figure}
\centering
\includegraphics[scale=0.38]{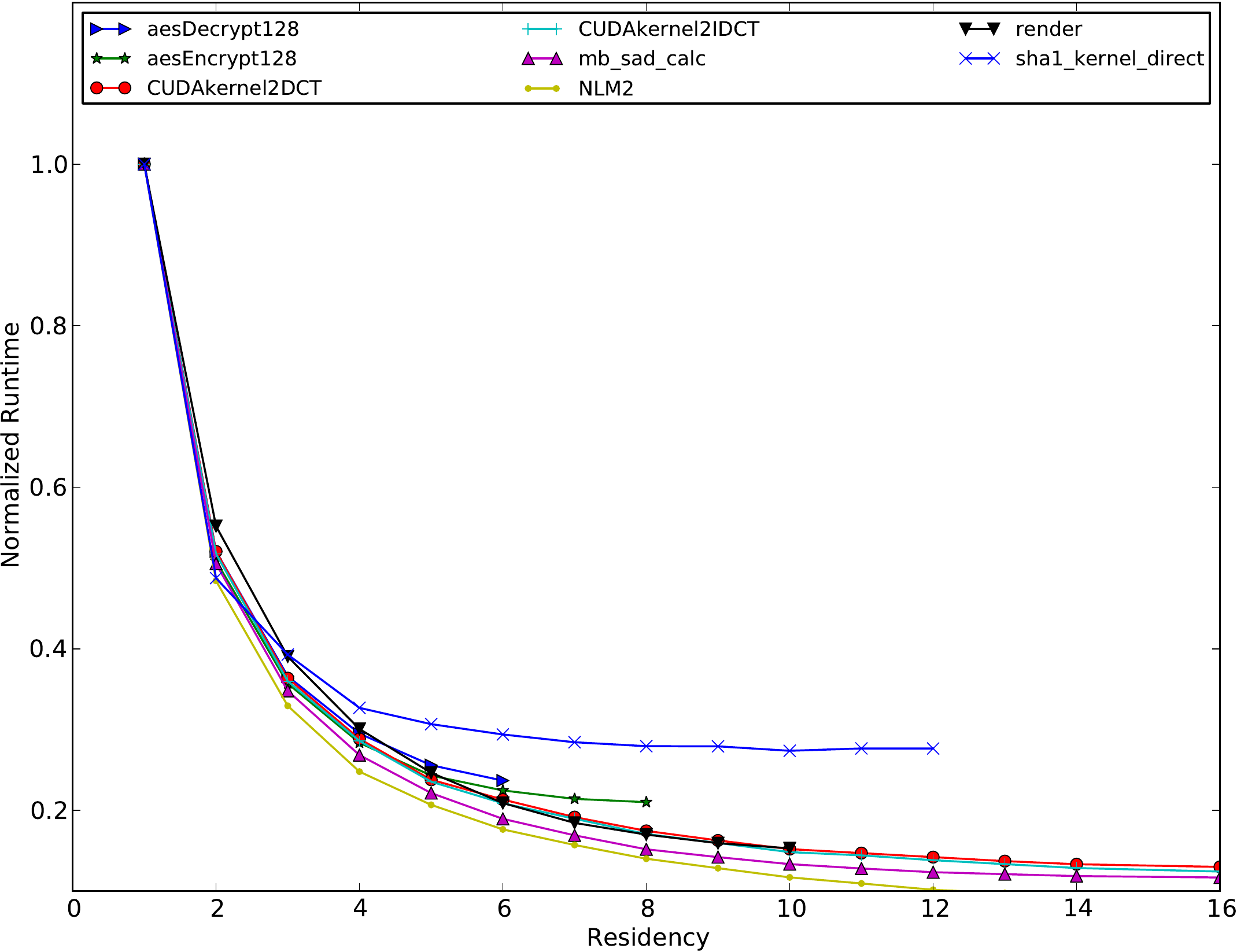}
\caption[Total kernel runtime at various residencies]{Total kernel runtime at various residencies normalized to runtime at residency 1. (Kepler equivalent of Figure~\ref{fig:runtimes_residency})}
\label{fig:kepler:runtimes_residency}
\end{figure}

Note that there is no data point for residency 15 in Figures~\ref{fig:kepler:tbtimes_residency} and~\ref{fig:kepler:runtimes_residency}.
Our method of controlling residency for a kernel on hardware involves changing the size of dynamic shared memory allocated to it during launch.
However, there is no size $x$ of shared memory that can be chosen such that $x$ is divisible by 256 and $15x \leq 49152$ and $16x > 49152$.
Here, 49152 is the total size of shared memory in bytes and 256 is the granularity at which it is allocated, also in bytes.

\begin{figure}
\centering
\includegraphics[scale=0.45]{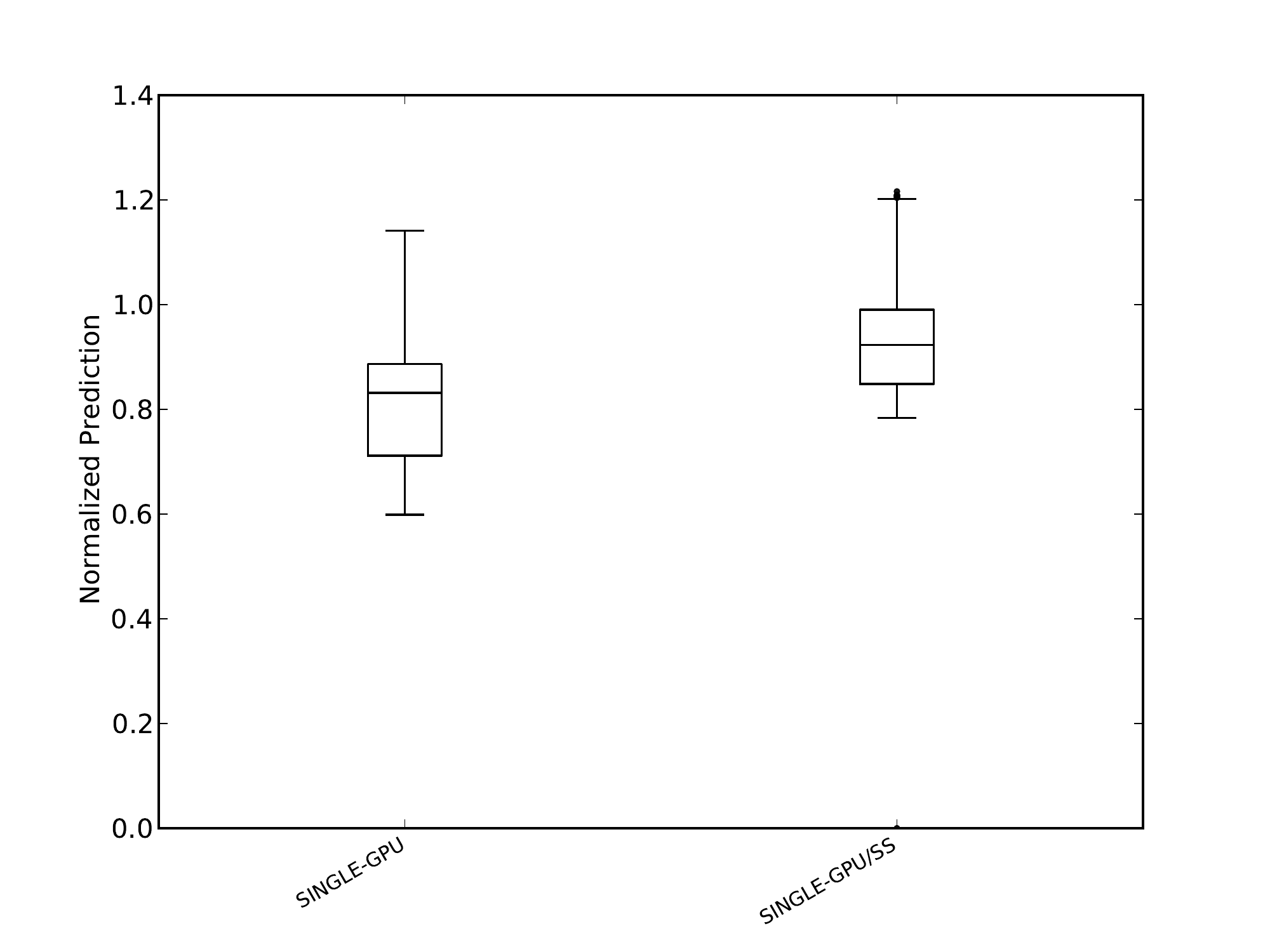}
\caption[Accuracy of the Simple Slicing (SS) Predictor]{Accuracy of the Simple Slicing (SS) Predictor. Runtime Predictions normalized to actual runtime. (Kepler equivalent of Figure~\ref{fig:pred_accu_results})}
\label{fig:kepler:pred_accu_results}
\end{figure}

Figure~\ref{fig:kepler:pred_accu_results} does not contain simulator results (i.e. simple-sim and mpmax).

\end{document}